\documentclass[12pt]{iopart}
\usepackage{epsfig}
\begin{document}
\jl{19}
\newcommand{\beq}{\begin{equation}}
\newcommand{\eeq}{\end{equation}}
\newcommand{\beqa}{\begin{eqnarray}}
\newcommand{\eeqa}{\end{eqnarray}}
   \def\esim{\mathrel{\rlap{\raise2pt\hbox{$\sim$}}
    \lower1pt\hbox{$-$}}}         
\def\lsim{\mathrel{\rlap{\lower4pt\hbox{\hskip1pt$\sim$}}
    \raise1pt\hbox{$<$}}}         
\def\gsim{\mathrel{\rlap{\lower4pt\hbox{\hskip1pt$\sim$}}
    \raise1pt\hbox{$>$}}}         

\newcommand{\bsg}{${\rm b}\to {\rm s}\gamma$}
\null
\vskip 3cm
\title{Non-Baryonic Dark Matter:\\
Observational Evidence and Detection Methods}
\author{Lars Bergstr\"om}
\address{Department of Physics,
Stockholm University,
         Box 6730, SE-113 85 Stockholm, Sweden, {\tt lbe@physto.se}}
\vskip .4cm
\begin{abstract}
The evidence for the existence of dark matter in the universe is reviewed.
A general picture
emerges, where both baryonic and non-baryonic dark matter is needed to explain
current observations. In particular, a wealth of
observational information points to the existence of a non-baryonic
component, contributing between around 20 and 40 percent of the critical mass density needed
to make the universe geometrically flat on large scales. In addition,
an even larger contribution from vacuum energy (or cosmological constant)
is indicated by recent observations. To the theoretically favoured particle
candidates for non-baryonic dark matter belong axions, supersymmetric particles, and
of less importance,
massive neutrinos.
The theoretical foundation and experimental situation
for each of these is reviewed. Direct and  indirect
methods for detection of supersymmetric dark matter are described in
some detail.
Present experiments are just
reaching the required sensitivity to discover or rule out some of these
candidates, and major improvements are planned over the coming years.
\end{abstract}


\submitted

\maketitle
\section{Introduction}

During the past few years, remarkable progress has been made in
cosmology, both observational and theoretical. One of
the outcomes of these rapid developments is the increased
confidence that most of the
energy density of the observable universe is of an unusual
form, i.e., not made up of the ordinary matter (baryons and electrons)
that we see around us in our everyday world.

The only realistic model for cosmology, and one that has become
more solidly established by recent observations, is the big bang model.
According to this Standard Model of cosmology, the universe
has been and still is expanding from a compressed and hot phase,
which existed some 15 billion years ago. The observational
cornerstones of
the big bang model -- the expansion of the universe, the fossil record
of light elements sythesized during the first few minutes, and the
existence still today of an intense thermal radiation field, the
cosmic microwave background -- created when
neutral atoms formed around 300 000 years after the big bang -- are more solid
than ever. In fact, cosmology is entering an era of precision measurements,
when information from all these three important processes is becoming
accurate enough that it can be used to clarify the detailed structure
and evolution of our Universe.

For example, big bang nucleosynthesis can be used
to determine the baryon fraction of the matter density in
the universe quite accurately
\cite{bbnreview},
and combined with analyses of galaxy cluster dynamics,
supernova data and the cosmic microwave
background radiation this gives  convincing arguments for
the existence of a large amount of non-luminous, i.e., dark, matter.
The matter content of the universe seems to be at least a factor of 5 higher
than the maximum amount of baryonic matter
implied by big bang nucleosynthesis. This dark matter is thus highly
likely to be ``exotic'', i.e, non-baryonic \cite{triangle}.

There are
also indications, although still somewhat preliminary, of the
existence of vacuum energy, corresponding to the
famous ``cosmological constant'' that Einstein introduced but
later rejected (although without very good reasons) in his
theory of general relativity. An alternative possibility
is a slowly varying scalar field, ``quintessence'' \cite{steinhardt}
which, like vacuum energy,  does not cluster at all (or
on very large scales only). This  possibility has recently been given
increased attention due to results from Type Ia supernova surveys
\cite{ariel,kirshner2} as will be discussed later in this review.

Although the existence of non-baryonic dark matter is now generally
accepted by most of the astrophysical community,
the nature of the dark matt\-er is one of the outstanding
questions in \mbox{standard} cosmology \cite{BGbook}.  In fact, since 1998
there is for the first time strong evidence for the existence
of non-baryonic dark matter in the universe, in the form of massive neutrinos.
This is due to the discovery of atmospheric neutrino oscillations in the
Super-Kamiokande experiment \cite{totsuka}. However, the natural
neutrino mass scale of around 0.1 eV which is implied by the neutrino
oscillation data
is not large enough to influence cosmology in a dramatic way (although
this type of non-baryonic dark matter would contribute about as much to
the total mass density of the universe as do the visible stars).

Even if the mass of one of the neutrinos is higher (up to, say, 5 eV
which would be possible if neutrinos are nearly degenerate in mass
or if there exist a fourth, sterile neutrino), the mass density would
not be large enough to explain the matter fraction of the cosmic average
density. There are also arguments from galactic structure against
an all-neutrino dark matter population. Therefore, it is natural to
ask which other fundamental particles could be good dark
matter candidates.

This review is organized as follows. In Section~\ref{sec:basics}
the general framework of modern cosmology is briefly reviewed.
In Section~\ref{sec:relic} the fundamental mechanism of thermal
production of relic particles in the early universe is explained.
In Section~\ref{sec:amount} various estimation methods for the
amount of different forms of mass and energy density on several
length scales in the universe are discussed. In Section~\ref{sec:baryons}
the possible contribution from hidden ordinary matter (baryons) is
treated, and the evidence for a population of dark baryonic
objects, ``MACHO''s (MAssive Compact Halo Objects), displayed.
The distribution of dark matter in
galaxies, in particular the in Milky Way which most of the detection
experiments are sensitive to, is discussed in
Section~\ref{sec:distribution}. Particle physics models for
non-baryonic dark matter are treated in Section~\ref{sec:models},
and the presently most discussed models for Weakly Interacting
Massive Particles (``WIMP''s), namely the
supersymmetric models, are reviewed in Section~\ref{sec:susy}.
These models, which have been developed in detail in the current
literature, also serve as useful templates for other, yet
unthought of, particle models for dark matter. Therefore,
Section~\ref{sec:detection} deals mainly with the methods of
detection of the lightest electrically neutral supersymmetric
particle,  the  neutralino, which is a quantum
mechanical mixture of the supersymmetric partners of the photon,
the $Z^0$ boson and the neutral scalar Higgs particles.
Section~\ref{sec:conclusions} ends the review with some concluding
remarks and an outlook.

\section{Basics of standard cosmology}\label{sec:basics}

The big bang model has its theoretical base in Einstein's general
theory of relativity, which has been thoroughly tested on many
different length scales, and which so far has agreed with all
observations.

The fundamental dynamical degrees of freedom of the gravitational
field are the components of the metric tensor, which enter
into the line element describing the distances  between nearby
points in space-time. To make cosmology tractable, one has to
assume that the universe is isotropic and homogeneous on the
average. This agrees, of course, with observations. On the
largest scales probed by the cosmic microwave background, the observed
temperature anisotropies, which are related to the density fluctuations
at the time of emission, are smaller than one part in $10^4$. (There is a dipole
component which is of the order of $10^{-3}$, but this is
interpreted as being of kinematical origin, caused by our motion with respect
to the local cosmic rest frame.)

One can show that every isotropic, homogeneous three-space
can be parametrized, perhaps after performing a coordinate
transformation, by coordinates which give a line element, squared
distance between points at a fixed time,  of
the form
  \beq
ds^2=a^2\left({dr^2\over 1-kr^2}+r^2d\Omega^2\right),\label{eq:spaceds}
 \eeq
where the constant $k$ after a suitable normalization of $r$ can
take one of the three values $k= -1,\,0,\,+1$, depending on whether the geometry is
open, flat or closed. For a given value of $k$,  this
defines a one-parameter family of similar spaces, where the scale
factor $a$ is the parameter. These are the Friedmann-Lemaitre-Robertson-Walker
(FLRW) models. For the simplest case $k=0$, we see that the metric
 describes ordinary  Euclidean three-space with the scale factor $a$ giving the
overall normalization of physical distances. Due to the observed expansion of the
universe, we know that the scale factor depends on time, $a=a(t)$ with
$\dot a/a>0$ at the present time $t=t_0$,
and a basic task for standard cosmology is to determine $k$ and
to compute $a(t)$, both in the past and in the future.

\subsection{The Friedmann equation}
The time dependence of the scale factor $a$ is given by the Friedmann
equation (which follows from one of the components of Einstein's
tensor equations):

\beq
\left({\dot a\over a}\right)^2+{k\over a^2}={8\pi G_N\over 3}\rho_{tot},
\label{eq:friedmann1}
\eeq
where $G_N$ is Newton's gravitation constant, and $\rho_{tot}$ is
the total average energy density of the universe. The latter gets
contributions at least from matter, radiation and perhaps vacuum
energy (cosmological constant),
\beq
\rho_{tot}=\rho_{m}+\rho_{r}+\rho_{\lambda}.
\eeq
In particle physics units, $\hbar=c=1$,
(which means that all
dimensionful quantities can be expressed in powers of some mass unit,
e.g., 1 GeV),
$G_N$ has the dimension of
the inverse of a squared mass, the Planck mass, which has the huge
value of $1.22\cdot 10^{19}$ GeV.  From the particle physics viewpoint,
this means that gravity is governed by some still unknown theory
at superhigh energy, whose low-energy limit is Einstein's general theory
of relativity. The prime candidate for such a fundamental theory is
string theory or one of its related versions, but this is an active,
on-going field of research.

The overall value of the scale factor is arbitrary, only relative changes
are measurable. The Hubble parameter (which depends on time) is
\beq
H(t) ={\dot a(t)\over a(t)}
\eeq
and governs the local expansion according to Hubble's law, $v=Hd$,
where $v$ is the recession velocity and $d$ is the physical distance.
Observationally, the present value $H_0$ of the Hubble parameter
(called the Hubble constant)
is uncertain to around 20 \%, and one usually writes
\beq
H_0=h\cdot 100\ {\rm km}\,{\rm s}^{-1}\,{\rm Mpc}^{-1}
\label{eq:hubblenow}
\eeq
with $h=0.65\pm 0.15$ encompassing most recent
observational determinations of the Hubble constant
(1 Mpc is approximately $3.1\cdot 10^{22}$ m). In SI units,
\beq
H_0=3.2\cdot10^{-18}h\ {\rm s}^{-1}.
\eeq
Consequently, the linear length scale of the universe is presently
being stretched by a fraction of
$3.2\cdot 10^{-18}h$ per second. The inverse of
this expansion rate defines a time, the Hubble time $t_H=9.8\ h^{-1}$
years. In standard cosmology, the present age of the universe is
of this order of magnitude modulo a numerical factor slightly less than
unity reflecting the fact that the expansion rate most of the time
should have been larger than today. In particle physics units
($\hbar=c=1$) the expansion rate has the dimension of mass, the
numerical value being $2.1\cdot 10^{-42}h$ GeV. The large value
of  the Planck mass compared to the expansion mass scale (and all other
other fundamental mass scales in nature) is one of the unsolved
problems of theoretical physics.

We see from Eq.~(\ref{eq:friedmann1}) that the universe is flat ($k=0$)
when the energy density equals the {\em critical density} $\rho_{c}$,
given by
\beq
\rho_{c}\equiv {3H^2\over 8\pi G_N}\label{eq:rhocrit}.
\eeq
From Eq.~(\ref{eq:hubblenow}), its present numerical value is
computed to be $\rho^0_{c}=1.9\cdot 10^{-38}h^2\ {\rm
kg\,m}^{-3}$.
The Friedmann equation (\ref{eq:friedmann1}) can then be
written in the equivalent form
\beq
{k\over H^2a^2}+1={\rho\over\left({3H^2\over 8\pi
G_N}\right)}\equiv\Omega,
\eeq
where
\beq
\Omega\equiv {\rho\over \rho_{c}}\label{eq:rhocrit2}
\eeq
or
\beq
{k\over H^2a^2}=\Omega-1 .\label{eq:timedep}
\eeq
We thus see that $\Omega$ is the energy density in units
of the critical density, and $\Omega=1$
is equivalent to having a flat universe ($k=0$).

The expansion of the universe means that the scale factor $a(t)$
has been increasing since the earliest times after the big bang.
The first observational evidence for this was of course Hubble's
detection of a cosmological redshift of the light emitted by distant
galaxies. For an emitted wavelength (e.g., associated with a specific
spectral line)
$\lambda_{emit}$ and an observed wavelength $\lambda_{obs}$
the redshift parameter $z$ is defined
by
\beq
1+z\equiv{\lambda_{obs}\over\lambda_{emit}}.\label{eq:redshift}
\eeq
In the standard FLRW model of cosmology, the redshift is
 related to the change in scale factor $a(t)$ through
the relation (see \cite{BGbook} for a simple derivation)
\beq
1+z={a(t_{obs})\over a(t_{emit})}.\label{eq:redFLRW}
\eeq

\section{Creation of relic particles}\label{sec:relic}

In a universe which expands, there is at each epoch a mixture of different
particles in thermal contact with each other, maintaining a
temperature which evolves with time. For instance, when the temperature
of the universe was much larger than a few MeV,
electromagnetic interactions kept electrons and positrons in thermal
equilibrium and weak interactions kept also neutrinos in equilibrium.
However, to maintain this equilibrium, interactions had to be frequent
enough. The critical time scale is set by the Hubble parameter, which has
the dimensions of 1/time. The
interaction rate per particle with velocity $v$
is given by $\Gamma=n\sigma v$, where
$n$ is the number density of ``target'' particles and $\sigma$ is
the cross section\footnote{In a more refined treatment, one has to solve
the Boltzmann transport equation, which couples all different particle
species, in a background metric given by the FLRW model. This also
allows to follow the deviations from exact thermal equilibrium
caused by the expansion. The results
of our simplified discussion are in rough agreement with this treatment.}.
Choosing units such that $\hbar=c=k_B=1$ (i.e., also the temperature
is measured in mass units),
the thermal equilibrium number density for a particle species
of mass $m_i$ is, at temperature $T$
\beq
n_i={g_i\over \left(2\pi\right)^3}\int f_i({\mathbf p})d^3{\mathbf p},
\label{eq:therm}
\eeq
with
\beq
f_i({\mathbf p})={1\over e^{{E_i\over T}}\pm 1},
\eeq
where $E_i=\sqrt{{\mathbf p}^2+m_i^2}$ is the energy and the plus
sign applies for fermions which obey Fermi-Dirac statistics, the minus sign for bosons
(Bose-Einstein statistics). In the non-relativistic
limit $T\ll m_i$, the integral can be solved, and one obtains the
familiar result
\beq
n_{NR} = g_i\left({m_iT\over 2\pi}\right)^{3/2}e^{-m_i/T}
\eeq
which shows the exponential suppression of heavy particles expected
due to the relativistic rest energy $E_i^0=m_i$. The degeneracy
parameter $g_i$ counts the independent degrees of freedom (i.e. the
number of helicity states). For photons, $m_\gamma=0$, $g_\gamma=2$.
For neutrinos, $m_\nu$ is unknown but presumably very small, with
a total of $g_\nu=6$
(three neutrinos of negative helicity and three antineutrinos of positive
helicity -- the other helicity states are missing reflecting the
maximal parity violation of the weak interactions). In the other
solvable regime, the relativistic one ($T\gg m_i$), the rest energy
of particles can be neglected, and the number density becomes
\beq
n_R=s_ig_i{\zeta(3) T^3\over \pi^2}
\eeq
where $s_i=1$ for bosons and $s_i=3/4$ for fermions.

The important point to notice is the absence of any exponential
suppression, which means that at a given epoch the number density
of relativistic particles by far outweighs that of nonrelativistic
particles, as long as thermal equlibrium is maintained.

The corresponding energy densities are also easily obtained by
inserting a factor $E_i$ in the integrand of Eq.~(\ref{eq:therm}).
Again, nonrelativistic particles contribute a negligible amount,
and for relativistic particles the result is
\beq
\rho_R=u_ig_i\left({\pi^2 T^4\over 30}\right)
\eeq
with $u_i=1$ for bosons and $u_i=7/8$ for fermions.

The Friedmann equation (\ref{eq:friedmann1}) now shows that
the expansion rate $H$ evolves as
\beq
H^2 = {8\pi G_N\over 3}{\pi^2 T^4\over 30}\sum_ig_iu_i ,
\eeq
where the small terms from curvature and non-relativistic
particles have been neglected.

We are now in a position to compute the temperature at which neutrinos
decoupled in the early universe. The cross section for neutrinos
is of weak interaction
strength meaning $\sigma\propto G_F^2s$ with $G_F$ the
Fermi coupling constant $\sim 1/m_W^2$, and $s$ the total energy squared
in the centre of mass, i.e., $s\propto T^2$ for relativistic particle
collisions. For thermal equilibrium to be maintained, one has to
demand that each particle has interacted at least once during a
Hubble time $\sim H^{-1}$.
Comparing thus the expansion rate $H$ with $n\sigma v$ (using that
$n\propto T^3$ and $v$ is of order unity),
neglecting the constant numerical factors, a simple calculation shows that
at a temperature of around 3 to 4 MeV the interaction rate fell
below the expansion rate. After that, neutrinos would have a fixed
comoving number density and a distribution that would look like thermal,
but the energies of neutrinos, while still relativistic, would
redshift with the expansion similarly to the photons of the microwave
background radiation. A more careful evaluation, keeping all
numerical factors, gives as a result that the ``neutrino background''
which should exist in the universe, now has a number density of
around 50 cm$^{-3}$ neutrinos per degree of freedom (i.e. of the
order of 300 cm$^{-3}$ in total).

We now see an important possibility, which was historically
the first serious attempt to explain the dark matter problem
using particle physics. If neutrinos have a finite but small
mass, they should  have the number density just computed,
but since the effective
temperature due to the expansion may be lower than
the neutrino mass they would be non-relativistic at the present
epoch and contribute
significantly to the energy density of the universe. Quantitatively,
one obtains
\beq
\Omega_\nu h^2={\sum_i m_{\nu_i}\over 93\ {\rm eV}}.\label{eq:relicneutrino}
\eeq

If the particle is more massive, so that it is non-relativistic at
decoupling, the mathematical treatment is a little more involved
(since the exact form of Eq.~(\ref{eq:therm}) has to be used when tracking
the interaction and expansion rates, and also the cross section
will be a more complicated function of temperature), but identical in principle.
A general framework for obtaining a numerical solution, even
in the presence of resonances and particle production thresholds in
the expression for the cross section,  can be
found in \cite{gelminigondolo}, and a simple treatment in \cite{roszkowski}.
As a first-order estimate,
the contribution
to $\Omega_M$ from a massive particle $\chi$ depends only on
its annihilation cross section $\sigma_A$ and not explicitly
on its mass (to logarithmic
accuracy). The resulting $\Omega_\chi$ is then \cite{jkg}
\beq
\Omega_\chi h^2\sim {3\cdot 10^{-27}\ {\rm cm}^3{\rm s}^{-1}\over \langle \sigma_A v\rangle} \label{eq:relicdensity}
\eeq
where $\langle\ldots\rangle$ indicates taking a thermal average.
Usually, $\sigma_A v$ is dominated by S wave processes, and then it
is nearly velocity-independent and therefore temperature-independent.
If the S wave is forbidden by selection rules, or if there exist
particle poles
or thesholds in or near the physical region for the process, the
temperature corrections can be large.
An interesting consequence of the result  in Eq.~(\ref{eq:relicdensity})
is that massive
particles which have
interactions of the order of the weak interactions naturally give
contributions to $\Omega_M$ of order unity. The generic
name for such a dark matter candidate is a WIMP (Weakly Interacting
Massive Particle).
For massive particles (heavier than a few GeV), the freeze-out temperature
(the temperature at which the number density of the particle leaves the
thermal equilibrium curve) is typically 4 to 5~\% of the particle mass,
i.e., they are non-relativistic already at the decoupling epoch.

Although thermal production of stable particles is a generic,
unavoidable
mechanism in the big bang scenario, there are several additional
processes possible.
For instance, some very heavy particles were perhaps never
in thermal equilibrium. Non-thermal
production may also occur
near cosmic strings and other defects as
happens, for instance, in the case of axions discussed in Section~\ref{subs:axions}.
Near the end of a period of early inflation, several mechanisms
related either to the inflaton field or to the strong gravity present
at that epoch could contribute to nonthermal
production (see, e.g., \cite{tkachev}).

\section{The Amount of Matter and Energy in the Universe}\label{sec:amount}

The big bang model contains a strong implicit connection between
particle physics and cosmology. In the early universe, the
thermal energies were high enough that even the heaviest of
the known particles could be pair created and destroyed in
near-equilibrium processes. As time evolved, the universe may
have gone through a whole series of phase transitions, e.g.,
the electroweak transition when the $W$ and $Z$ bosons got their
mass, and the confinement transition from a quark-gluon plasma
to bound quark (and antiquark) states.

\subsection{Inflation}\label{subs:inflation}

The particle physics -- cosmology connection became much
stronger in the beginning of the 1980s with the discovery of
the cosmological inflation mechanism \cite{inflation}.
Traditional big bang cosmology suffers
from problems concerning initial conditions. For instance,
why is it that the microwave background radiation has nearly
the same temperature even in oppositely directed regions on
the sky which
cannot have been in causal contact at the time of emission?
This is sometimes called the horizon problem.
Inflation
solves this by assuming that some scalar ``inflaton'' field in the early
universe slowly changed its expectation value due
to a temperature-dependent effective potential which left the field
temporarily at a non-minimum. If the potential
energy density during this ``slow-roll'' phase was almost constant,
and dominated the energy density of the universe at that time,
it acted in a simliar way as a true
 cosmological constant. This would cause according to Eq.~(\ref{eq:friedmann1})
 a constant value of the Hubble parameter $H$, and the equation
 \beq
\left({\dot a\over a}\right)^2=H^2={\rm const}
\label{eq:inflation}
 \eeq
 is easily solved to give
 \beq
a(t)\propto e^{Ht}
\eeq
 which means a superluminal
(exponential) expansion\footnote{This is in no contradiction with
relativity, since it is space itself  which expands. Locally,
in any given small region,
matter has to move with a velocity less than the velocity
of light.}. This enormously rapid expansion
implied that regions that had been in
causal contact  moved outside the horizon of each other.
The inflationary epoch stopped when the scalar field had rolled
down to the true minimum of the potential.
Due to the energy released by damped,
coherent oscillations of the field around
this minimum the universe was refilled
with particles and radiation and an enormous amount of
entropy was created (the reheating epoch). After this,
the universe
restarted in something similar to the traditional big bang
cosmology, but with very smooth and flat initial conditions.

In particular, a generic prediction from inflation is thus that
the universe is very close to flat, i.e., $k=0$.
Then
$\Omega_{\rm tot}=\Omega_0=1$, with the subscript 0 referring to the value at
 the present time $t=t_0$,
 $\Omega_0\equiv \Omega_{\rm tot}(t_0)$.

Inflation is strictly speaking not a mandatory ingredient in
the modern Standard Model of cosmology, but in lack of alternative
explanations of, e.g., the horizon problem it has a
prominent place in theoretical cosmology. An additional virtue
of inflation is that it gives a possible explanation of
the origin of the small primordial fluctuations seen in
the microwave background. This could be caused by vacuum
fluctuations of the inflaton field, which are expected
in most models of inflation to be gaussian and nearly
scale-invariant. As more accurate measurements
of the cosmic microwave background are soon to be expected
(see Section~\ref{subs:cmbr}), these features will be tested
further.

\subsection{Different forms of energy density}
 In general, if $\Omega\neq 1$ it is  time-dependent, see Eq.~(\ref{eq:timedep}).
 Even if $\Omega_0=1$ as predicted by inflation,
 the various different contributions to $\Omega_{tot}(t)$ depend
 differently on time, and therefore on redshift.
It is  thus important to determine how the total energy density in the universe
 is divided into its main
components $\Omega_M=\Omega_B+\Omega_{DM}$ and $\Omega_\Lambda$,
with $\Omega_0=\Omega_M+\Omega_\Lambda$. Here we have separated the
mass contribution into a baryonic part $\Omega_B$ and a non-baryonic,
dark-matter part $\Omega_{DM}$. We take the capital subscripts to
refer to present-day values, i.e., $\Omega_M=\Omega_m(t_0)$ etc. From
Eq.~(\ref{eq:timedep}) it can be seen that an eventual deviation from
$\Omega_0=1$ can be interpreted as a present-day contribution from
curvature to the expansion rate. Sometimes, one therefore introduces
\beq
\Omega_K={-k\over a_0^2H_0^2}
\eeq
in which case a sum rule is obtained:
\beq
\Omega_M + \Omega_\Lambda + \Omega_K =1.
\eeq
This makes the three $\Omega$ variables
similar to the so-called Dalitz  variables in particle
physics, and the various portions of this parameter space
allowed or disfavoured by
measurements may be displayed in triangular ``Dalitz plots'' (see,
e.g., \cite{triangle}).

Even if $\Omega_0=\Omega_M+\Omega_\Lambda=1$, the relative weights of
the various contributions
generally depend on time
and therefore on redshift $z$. In particular, the matter density
scales as $(1+z)^3$ since a constant comoving number density causes
the physical mass density  to be diluted with the changing volume, whereas
the cosmological constant is simply constant. This means that
\beq
{\Omega_\lambda\over\Omega_m}= {\Omega_\Lambda\over\Omega_M \left(1+z\right)^3}
\eeq
so that  the
cosmological constant was insignificant at early times (high $z$),
but will eventually dominate the total density. At redshifts higher
than 1000, also radiation, scaling as $(1+z)^4$ was important,
but can be safely neglected in today's universe, since
measurements of the flux of cosmic background radiation photons which
dominate the radiation energy density give $\Omega_R\sim
10^{-5}-10^{-4}$. Of course, there could in principle also exist
other forms of energy. If an equation of state relating the
energy density and the pressure of component $X$, $\rho_X=\alpha_Xp_X$
is valid, then $\Omega_X$ scales with $z$ as $(1+z)^{3(1+\alpha_X)}$.
Vacuum energy, or equivalently a cosmological constant, can then be seen as
a special case with $\alpha_\Lambda=-1$.
A general expression for the expansion rate is \cite{BGbook}
\beqa
{H^2(z)\over H_0^2}=\left[
\Omega_X\left(1+z\right)^{3(1+\alpha_X)}
+\Omega_K\left(1+z\right)^2\right. \nonumber \\
\left. +\Omega_M\left(1+z\right)^3
+\Omega_R\left(1+z\right)^4
\right].\label{eq:general}
\eeqa
The different dependence on $z$ is the key to disentangling all
the possible contributions to $\Omega_0$,  by observations
at different, and preferably large, redshifts. Also, as we will see, planned precision
measurements of the cosmic microwave background radiation may be very
helpful in determining all these parameters, as well as several other
presently ill-determined cosmological parameters.
\subsection{Supernova cosmology}

Until a couple of years ago, the theoretically favoured FLRW model was
the Einstein -- de Sitter model which has $\Omega_M=1$, $\Omega_\Lambda=0$.
The justification for this model is that it is compatible with inflation,
and avoids the severe finetuning problem of having a cosmological constant
$\Lambda$ which is non-zero but still some 120 orders of magnitude smaller than
inferred from the natural energy scale of quantum gravity,
$\Lambda\sim m_{\rm Pl}^4$ \cite{weinberg,dolgov}.
It is, however,
impossible to quantify how severe this finetuning is since there
exists at present no established theory of quantum gravity. Also, most
dynamical estimates of $\Omega_M$ on galaxy and cluster scales
have consistently given much smaller values of $\Omega_M\sim 0.2-0.4$
\cite{peeblesbook}.

Since agreement between observations and the big bang nucleosynthesis
(BBN) prediction of helium, lithium and deuterium abundance and
puts an upper limit to
 the  baryonic contribution  $\Omega_{B}$
of  \cite{bbnreview}
\beq
\Omega_{B}h^2\leq 0.019,\label{eq:bbn}
\eeq
non-baryonic dark matter has to dominate the energy density by a large
factor, even if $\Omega_M$ would be as low as 0.3. Also, we will see
that there are other problems with an all-baryon universe especially
as concerns the formation of structure.

\begin{figure}[!htb]\begin{center}
\epsfig{file=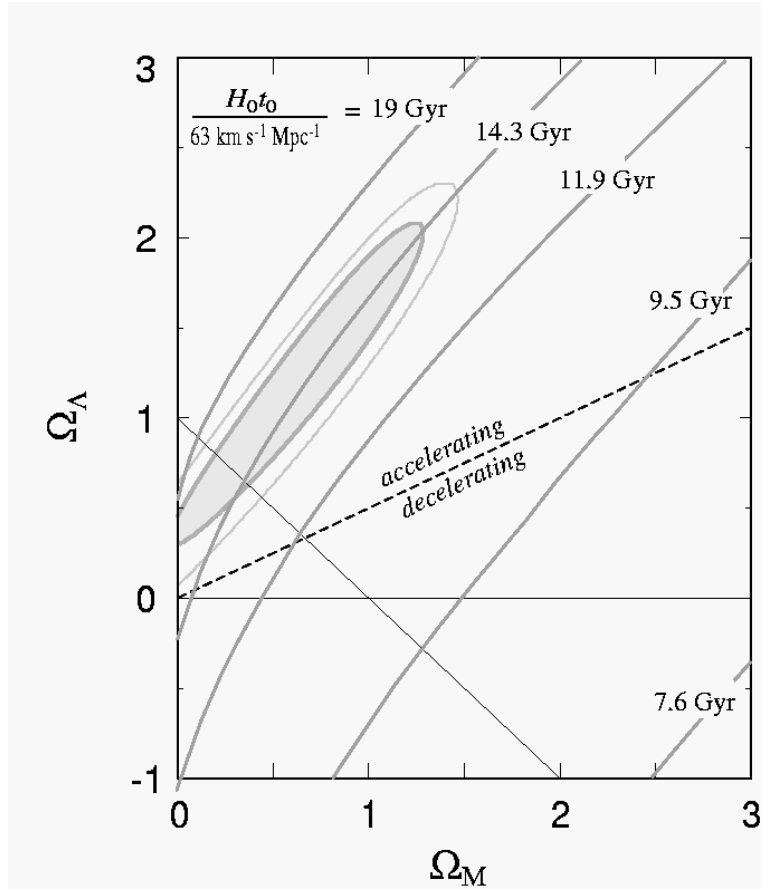,width=10cm}
\end{center}
\caption{Best-fit coincidence regions in the
$\Omega_{\rm M}$--$\Omega_\Lambda$ plane from the analysis of the
{\em Supernova Cosmology Project} \protect\cite{ariel}. The dark and
light ellipses show the 68 per cent and 90 per cent confidence regions. A flat
Universe ($\Omega_{\rm K}=0$) would fall on top of the diagonal
solid line passing through the prediction from inflation
$\Omega_0=\Omega_M+\Omega_\Lambda=1$. To the right
 of that line the Universe is closed, and to the left it is open.
The dashed line
shows the division between acceleration and deceleration for the
Universe. Also shown are isochrones of constant age.
Figure kindly provided by A. Goobar,
The Supernova Cosmology Project \protect\cite{ariel}. }
\label{fig:sn}
\end{figure}

A recent shift of most favoured cosmological model has occurred
thanks to the new determinations of the geometry of the universe
using Type Ia supernovae as standard candles. The idea is simply that the
line element in Eq.~(\ref{eq:spaceds}) implies a different ratio of
angular diameter to radial distance for the different values
of $k$. Moreover, since the finite propagation speed of light
means that one probes an earlier epoch of the universe when
making observations at large redshifts, the different $z$-dependence
of the various contributions
to $\Omega$ will  influence the relation between observed luminosity
and the redshift (see Eq.~(\ref{eq:general})).
This new type of analysis of the energy contents of the universe
seems to require a non-vanishing $\Omega_\Lambda$ \cite{triangle}, especially
when combining the supernova data with data from the cosmic microwave
background (CMB)
radiation \cite{lineweaver}. The
presently ``best-fit'' cosmological model which is compatible with $\Omega_0=1$
has $\Omega_M\sim 0.3\pm 0.2$,
$\Omega_\Lambda\sim 0.7\mp 0.2$, where the error estimates are
indicative only. Relaxing the requirement $\Omega_0=1$ gives a
larger, cigar-shaped region in the $\Omega_M$--$\Omega_\Lambda$ plane
(see Fig.~\ref{fig:sn}), but
data from both groups \cite{ariel,kirshner2}
 support a non-zero $\Omega_\Lambda$ at the level of two or three
standard deviations. Fixing $\Omega_{0}=1$, as predicted by inflation
and as seems preferred by the cosmic microwave background data discussed
in Section~\ref{subs:cmbr}, the mentioned values
$\Omega_\Lambda\sim 0.7\pm 0.2$, $\Omega_M\sim 0.3\mp 0.2$ emerge.

This is a field
of great potential which evolves rapidly, and the use of the
infrared camera on the Hubble Space Telescope should make follow-up
observations of high-$z$ objects easier. Eventually, a dedicated
satellite would allow more than an order of magnitude improvement
of the statistics with an accompanying hope of a better understanding
the sources of systematic errors, for instance evolution of metallicity
of the progenitor stars and the effects of absorption of grey dust.

\subsection{Gravitational lensing}

According to the predictions of Einstein's general relativity, the curvature
of space-time caused by matter gives rise to a deflection
of light rays, something first verified by Eddington in 1919.
Since the deflection angle is proportional to the mass of the
object causing the deflection (the ``lens''), one has in principle a good tool
for estimating directly the mass of astrophysical objects,
from planets and upwards to galaxies and galaxy clusters. In the
latter case, distant bright galaxies and galaxy nuclei
 such as quasars
are very useful as sources. Usually they can be considered as
being pointlike, and an excellent signature of gravitational
lensing is the appearance of multiple images of one and the same
quasar. The masses of the lensing objects determine the angular
separation of images, and the frequency of lensing events
has a strong dependence on the geometry of the universe, in
particular the presence of vacuum energy \cite{fukugita,schneiderbook}.

The gravitational lensing analysis of \cite{kochanek},
based on the frequency of double images in large
surveys of quasars indicates that
there is plenty of dark matter; the 95~\%~c.l. limits are
$\Omega_{M}>0.38$ and $\Omega_{\Lambda}<0.66$. An analysis of the
number of arcs from gravitational lensing of clusters expected in
various cosmologies on the other hand
gives consistency for an open model with
$\Omega\sim 0.3-0.4$, but failure for closed models with or without
a cosmological constant \cite{bartelmann}. This consequently
may be marginally in conflict with the supernova results.
However, gravitational lensing
has its own problems with systematic errors related to selection
effects and uncertainties in the lens models. That individual galaxies
and galaxy clusters are completely dominated by dark matter with
the visible baryonic matter being subdominant is, however, demonstrated
without doubt in analyses of strong lensing of background
galaxies \cite{tyson}.

There is hope that future more complete surveys of gravitational lenses,
also the lensing of supernovas, may give information on whether
the dark matter is distributed in the form of compact massive objects
or in more diffuse form \cite{holz,metcalf,bggm}.

Although the analyses mentioned so far seem to point to a matter
 density smaller than critical,
one should not disregard the possibility
that the simple Einstein -- de Sitter model could be correct. For
instance, the supernova data  may  be plagued by larger than
anticipated systematical errors. Also,
there exist indications for a large value of $\Omega_M$ and a
small $\Omega_\Lambda$ on the very largest scales \cite{dekel}.

A problem for high-$\Omega_M$ models without cosmological constant was
until very recently the difficulty of reconciling the age of the
universe $t_{U}$ based on the present expansion rate $H_{0}$
with the estimated age of the oldest globular clusters. The values
$\Omega_{M}=1$, $\Omega_{\Lambda}=0$  give
$t_{U}=2/(3H_{0})$, which for $h=0.6$ implies $t_{U}\sim 10-11$ Gyr. The
determination of globular cluster ages on the other hand  used to give
$14-15$ Gyr as best estimates. Besides some doubts that may still remain
about the accuracy of these latter very indirect means of bounding
the age of the universe,  the recalibration of the
distance scale provided by the Hipparcos satellite  parallax
measurements has brought the globular cluster age limit down by 2-3 Gyr
\cite{chaboyer}, with the one-sided 95~\%~c.l. lower limit being 9.5
Gyr.
This means that  a critical universe is now marginally allowed without
a cosmological constant, if $h\lsim 0.6$, a value that is not far from
the current best estimates. The presence of a cosmological constant
generally increases the age for a given value of the Hubble constant,
so the ``age problem'' is not regarded as very severe anymore.

At present the emerging, but cautious, consensus is
that a non-zero cosmological constant may be needed. This recent paradigm
shift is mainly driven by the outcome of new observations. Still there
are severe theoretical problems related to the smallness
of the required cosmological constant expressed in natural
particle physics units. One cause of vacuum energy according
to field theoretical models is the expectation value of various
scalar field potentials. Since the mass scale of these vacuum expectation
values should be of the order of the mass scale of the physics involved,
a phenomenal cancellation between several sources of
has to occur, since the sub-eV value needed is many orders
of magnitude below any known phase transition (electroweak, QCD,
supersymmetric\ldots) that could be relevant
\cite{weinberg,dolgov}. On the other hand, no compelling theoretical
reason for the cosmological constant to be zero has been
found either. Schemes have been
invented where the vacuum
expectation value of a scalar field (``quintessence'') varies which
time so as to track the matter energy \cite{steinhardt}. It may
be possible to distinguish between this possibility and
a true cosmological constant with future
high-precision supernova and/or CMB data.

\subsection{The cosmic microwave background}\label{subs:cmbr}
Perhaps the strongest argument in favour of non-baryonic dark matter
comes from structure formation and the cosmic microwave background. As
already mentioned, the high degree of isotropy and the
nearly gaussian and scale-invariant nature of the anisotropy of the
CMB gives credence to inflation which in turn
generically predicts $\Omega_0=1$. In addition, it has
proven to be very difficult to bridge the epoch of
the emission the CMB (some 300 000 years after the big bang) and that of
formation of large scale structure in the universe without
the help of non-baryonic matter.

The
basic picture is simple: The COBE satellite observations of the
anisotropy of the
microwave background at the level of a few times $10^{-5}$ gives
the normalization of the density perturbations in the universe
at that epoch, at a redshift of around $z\sim 1100$ when the universe
quite suddenly became transparent due to the recombination of electrons
with hydrogen and helium nuclei.
According to the standard theory of structure formation in the
expanding universe, the gravitational instability caused an
adiabatic growth of these primordial fluctuations with time. However, when
the fluctuations were small, this growth was
only linear in the scale factor $a$, and could start
only when the universe was matter-dominated. With baryons only,
this occurred  roughly at the  time when the CMB radiation
was emitted. This means that
it is very difficult to understand how the highly non-linear
structures observed today could exist, since they have evolved much
more than the
factor 1100 that linear growth could yield.  This schematic picture
is verified by large numerical simulations of structure formation.

Non-baryonic dark matter helps since if it exists, matter domination
occured earlier, causing also the perturbations to start to
grow earlier. When
the radiation pressure on the baryons was released after the CMB epoch,
the baryons could fall into the gravitational wells already formed
by the non-baryonic dark matter.
In fact, the nice agreement between the COBE observations and
the predictions from inflation of an approximately
 gaussian scale-invariant spectrum of primordial perturbations,
may also be taken as a piece of evidence in favour of inflation which
predicts the presence of such
fluctuations caused by quantum effects during inflation.

Recently, there has
been a flurry of balloon and ground-based CMB experiments on smaller
angular scales, which probe the interesting dynamics of the acoustic
peaks in the primordial cosmic fluid \cite{marc}. Although one has to
await longer duration balloon flights and the MAP and Planck
satellite missions for precision
measurements, it seems that the recent data
favour a  critical universe of $\Omega_0 =1$
over an open universe of, say, $\Omega_0=0.3$ \cite{lineweaver,boomerang}.
A fortunate fact is that the location of the first acoustic peak
in the power spectrum of the CMB depends on a combination of
$\Omega_M$ and $\Omega_\Lambda$ which is orthogonal to that
probed by the deep supernova surveys. Therefore, a combination
of results from the two types of measurements gives much more powerful
constraints than each one separately. (See, e.g., \cite{lineweaver,perl}, where
a combined analysis  favours the set $\Omega_M\sim
0.4\pm 0.1$, $\Omega_\Lambda\sim 0.6\pm 0.1$.)

\subsection{Large scale structure}

In structure formation not only $\Omega_{DM}$ is important, but also
the type of particle making up the dark matter. If the particle is
very light (e.g., a massive but light neutrino),
it will be relativistic at the time structure starts to form
and will free-stream out of galaxy-sized overdense regions, so that
only very large structures can form early. The terminology \cite{szalay}
is such that this type of particle constitutes
 {\em hot dark matter} (HDM), and
structure forms top-down by the fragmentation of large
structures (``pancakes'') into smaller. This behaviour is
nowadays strongly
disfavoured
in view of observations of the distribution of galaxies at very high
redshift, but a hot dark matter component at, say, the 10 \% level
cannot be excluded.

Massive particles (GeV or heavier) will typically move with non-relativistic
velocities when they decoupled and can therefore
clump also on smaller scales. This
is {\em cold dark matter} (CDM), one important example being supersymmetric
neutral massive particle discussed later. In cold dark matter
scenarios, structure typically forms in a hierarchical fashion,
with small clumps merging in larger ones, forming galaxy halos and
successively larger structures.

Inbetween hot and cold dark matter there may exist {\it warm dark matter}
(WDM),
which could be made up of
keV scale neutral particles (e.g., the supersymmetric partner of
the graviton, the gravitino, in some models of supersymmetry
breaking). There the inverse of the mass scale of the particle defines a
length scale of structure formation below which early structure
is suppressed. Warm dark matter is
not particularly in favour
at the moment, both for particle physics and structure formation
reasons, but the possibility should be kept in mind.

There are particles, like the axion, which behave like cold
dark matter although they most likely are very light. This
is due to the nonthermal way they were produced.

 Although the details of the structure formation history of the universe
probably will remain unclear
 until the next generation of microwave background measurements and
 digital sky surveys are available, the formerly (in the 1980's)
so popular ``Standard Cold Dark Matter''
 (SCDM) model with $\Omega_{CDM}=0.95$, $\Omega_{B}=0.05$,
 the slope parameter of the scale invariant primordial power
 spectrum $n=1$ , and $h=0.5$  is now
 disfavoured
as observations have become more refined.
 The main problem is that normalization to the COBE spectrum at the
 largest scales causes by approximately a factor of 2  too much power on the
 smaller scales probed by galaxy and cluster surveys.
 However, with only small modifications such as adding an HDM
 component, tilting the primordial spectrum to $0.8-0.9$, decreasing
 $h$ below $0.5$ or a combination thereof one can get a quite satisfactory
 description of the data \cite{gawiser}.

A recent important piece of evidence in favour of the existence
of cold dark matter, with an estimated $\Omega_{M}\sim 0.3$ (for a flat
universe) or $0.5$ (for an open universe)
comes from observations of the Lyman-$\alpha$ forest \cite{croft},
combined with the observed mass function of galaxy clusters,
clearly necessitating a substantial amount of non-baryonic dark matter.
Combining data from cluster abundances, the CMB and the IRAS infrared
galaxy catalog, a range $0.30 < \Omega_M < 0.43$ can be inferred \cite{bridle}.

Still on very large scales, analyses of the peculiar velocity ``flow''
of large clusters and other structures seem to need a lot of
gravitating matter for its explanation, at least $\Omega_M > 0.4$ \cite{dekel}.
The peculiar velocity field obeys the equation
\beq
\nabla\cdot {\bf v}={-\Omega_M^{0.6}\over
b}\left({\rho-\langle\rho\rangle\over \langle\rho\rangle}\right),
\eeq
where $b=\delta\rho_{Gal}/\delta\rho_{M}$ is the ``biasing'' parameter
which tells how light traces mass. The combination $\Omega_M^{0.6}/b$ is
determined by the analysis of \cite{dekel} to be $0.89\pm 0.12$, which
is  consistent with $\Omega_M=1$, $b=1$. Using the theoretical limit
$b>0.75$, a 95~\% c.l. limit of $\Omega_M>0.33$ can be given.

\subsection{Galaxy clusters}

The analysis of galaxy clusters is starting to converge to a universal
value of $\Omega_M$. It is true that here are some
indications \cite{blanchard} from
the temperature-luminosity relation for rich clusters that a high
value ($\Omega_M\sim 1$) might be needed. On the other hand
\cite{nbahcall,sasha,carlberg} most dynamical estimates
are more consistent
with a lower value, $\Omega_M\sim 0.2-0.3$ on cluster scales.

Of special importance is the large amount of X-ray
emitting hot gas
present in
 rich galaxy clusters. These systems are large enough that the baryon
fraction should be a good tracer of $r_{B}=\Omega_{B}/\Omega_{M}$.
Estimates \cite{clusterfraction} give a value of $r_{B}\sim 0.1 -
0.2$,
which combined with the BBN determination of $\Omega_{B}$ gives
$\Omega_{M}\sim 0.2$ if the high deuterium measurement \cite{highd}
is correct,
$\Omega_{M}\sim 0.5$ for the low deuterium abundance case
\cite{lowd}, with probably rather large systematic uncertainties
related to the limited understanding of how clusters formed and
how the gas thermalized.

Also the microwave background can be
used to extract $\Omega_M$, namely through the Sunyayev-Zel'dovich
(SZ) effect, by which the CMB gets spectrally distorted through
Compton scattering on hot electrons in galaxy clusters. This effect
of course depends on the baryonic fraction of the cluster and
could therefore give an estimate of $\Omega_M$ with the same assumption that
the baryon fraction of clusters is a fair sampling of the
cosmological baryon fraction. With  present SZ data, a value
of $\Omega_M h\sim 0.25$ is obtained \cite{carlstrom}.

It is
interesting that cluster mass estimates based on gravitational lensing,
X-ray emission, the SZ effect
 and galaxy motions all give similar mass estimates
within about a factor of two.

\subsection{The need for non-baryonic dark matter}
\begin{figure}[!htb]\begin{center}
\epsfig{file=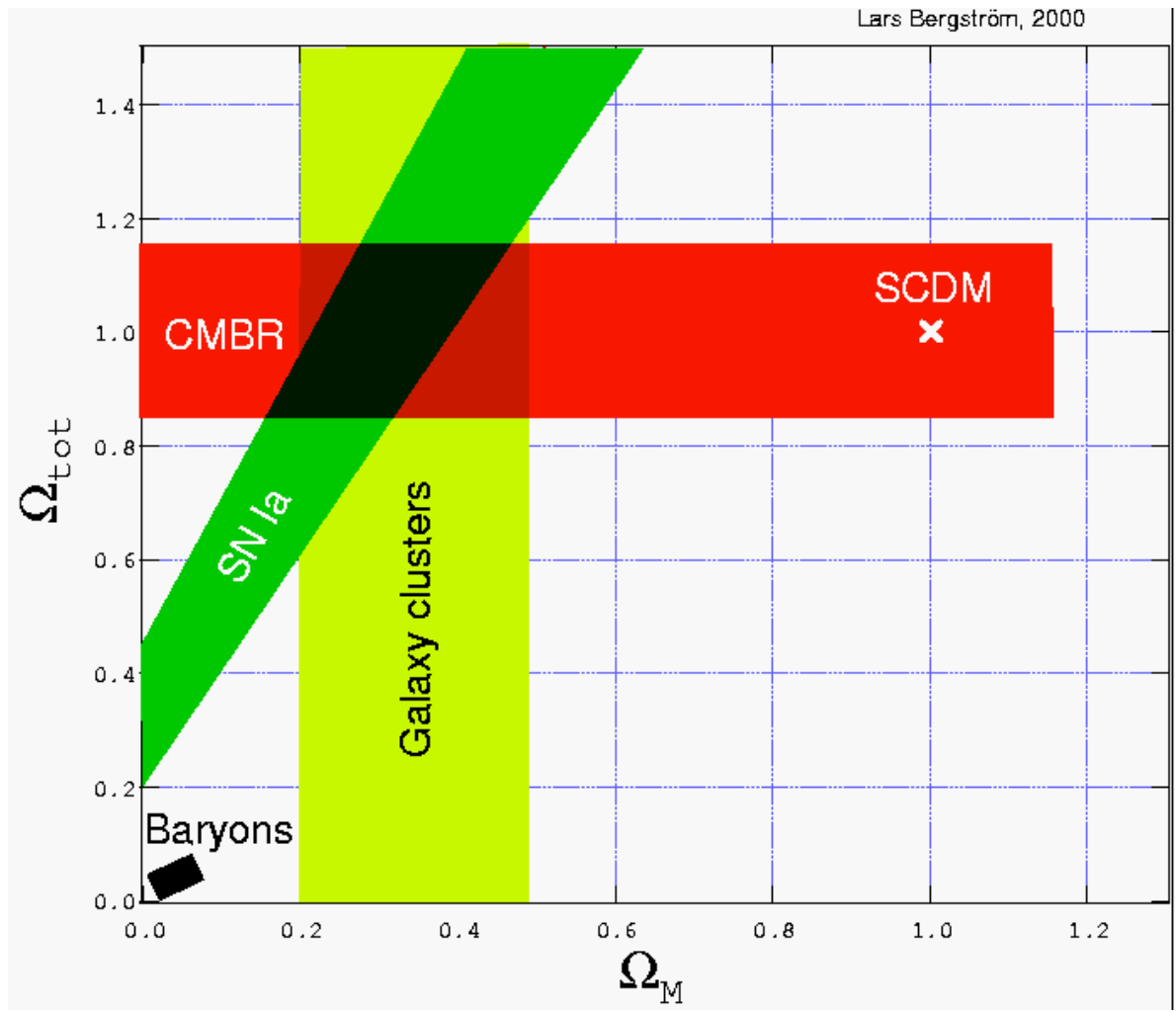,width=\textwidth}
\end{center}
\caption{Approximate best-fit coincidence regions in the
$\Omega_{\rm M}$--$\Omega_{\rm tot}$ plane (assuming that only
$\Omega_M$ and $\Omega_\Lambda$ contribute to $\Omega_{\rm tot}$),
from the joint analysis of
Type Ia supernovae \protect\cite{ariel} (diagonal shaded band),
cosmic microwave background radiation \protect\cite{boomerang}
(horizontal band) and X-ray cluster mass measurements
\protect\cite{clusterfraction}
(vertical band). As can be seen, there is a concordance region for
$\Omega_M\sim 0.3$, $\Omega_{\rm tot}\sim 1$ (i.e., $\Omega_\Lambda\sim 0.7$).
Also shown is a region at very small $\Omega_M$ which bounds the baryons-only
contribution according to big bang nucleosynthesis
\protect\cite{bbnreview}. The formerly
popular ``Standard CDM'' model, SCDM, is shown as the cross at
$\Omega_{\rm tot}=\Omega_M=1$.
}
\label{fig:omega}
\end{figure}

To summarize at this point: A variety of independent estimates of the
matter density in the universe point to a value larger than the
maximal value provided by baryons alone according to nucleosynthesis.
The need for nonbaryonic dark matter is therefore striking. If the
``natural'' theoretical prediction $\Omega_{M}=1$,
$\Omega_{\Lambda}=0$ is fulfilled is a different question, for which
most of the observational data today do not give support, except
some weak indications at the largest scales. If one accepts, as seems
implied by the deep supernova surveys, a non-vanishing cosmological
constant $\Lambda$, then a satisfactory description of most
cosmological observations is obtained by a ``$\Lambda$CDM model'' with
$\Omega_B\sim 0.05$, $\Omega_{\rm CDM}\sim 0.25$, $\Omega_\Lambda\sim 0.7$,
see Fig.\ref{fig:omega}.

\subsection{Problems with Cold Dark Matter}

There is thus compelling evidence for dark matter on
large scales in the universe, and big bang nucleosynthesis
forbids the larger part of it to be baryonic. The present-day
``best fit'' model is a flat $\Lambda$CDM
model with $h\sim 0.65$, $\Omega_M\sim 0.3$
and $\Omega_\Lambda\sim 0.7$, where baryons only contribute $\Omega_B\sim 0.05$
to $\Omega_M$, the rest being due to cold, non-baryonic dark matter.
Although this model fits
most of the cosmological data remarkably well,
there are some reasons for caution.

One potential problem for the $\Lambda$CDM model, as for most other
versions of CDM models, is that it predicts galaxy halos that
have a steeper density profile in the inner parts than may be observed
\cite{moore94,flores}. Also, N-body simulations seem to
give much more substructure of dark matter within clusters and galactic
halos than what is observed, e.g., in terms of the number of dwarf satellite
galaxies to the Milky Way. It is not clear, however, how serious these
problems are in view of the large uncertainties present in the N-body
simulations of structure formation as regards, e.g., the shape and slope
of the primoridial spectrum of initial perturbations \cite{kamlid}. The
situation concerning these numerical simulations is presently confused,
with different groups reaching different conclusions concerning, e.g.,
the dark matter density profiles near the centres of galaxies. This
is a problem we will return to when discussing experimental signatures
of various cold dark matter particle candidates. It has recently
been proposed \cite{spergel}, that the problem can be circumvented
if the dark matter is self-interacting with rather large cross section
but interacts weakly with ordinary matter and also has a suppressed
annihilation rate. It remains to be seen however, if a realistic
particle physics scenario can be found for this type of model.

A major observational difficulty is to extract the rotation curve
in the inner parts of galaxies. One recent analysis \cite{bosch} finds, for
instance, that if one takes beam smearing into account the
density profiles of low surface brightness galaxies are indeed
consistent with having a steep central cusp, in contrast to
previous claims. This is a field where more refined observations
are clearly needed.

The introduction of the cosmological constant adds a new parameter
which of course makes a fit to some sets of
cosmological data easier (and it has to
be reminded that most of the data may have large systematic uncertainties).
From a theoretical point of view, other modifications of the parameters
may then be preferrable, such as, for example, lowering $h$ to well below 0.5
\cite{turner_lowh}. This does not seem easy to do given the
concurrence of several methods which give the ``canonical'' range of
$0.65\pm 0.15$. The point made here, however, is that there is no simple
cosmological model which does not have problem with at least some sets
of data. In that situation it may be wise to keep several
possibilities open, while awaiting data of higher precision.

\section{Baryonic Dark Matter}\label{sec:baryons}
\subsection{Big bang nucleosynthesis}
The BBN limit in Eq.~(\ref{eq:bbn}) is important, since it implies that if
observations give a value of the total energy density above the BBN
value, non-baryonic dark matter has to be present (or baryons have to
be  hidden
in some non-standard way at the time of nucleosynthesis), even if the
total $\Omega_M$ turns out to be less than unity. Indeed, we have
seen that there are several
independent indications that $\Omega_M > 0.1$ (and hardly any estimates
at all that fall below that limit). Of course, it has long been
recognized that even  the minimum value of $\Omega_{B}$
allowed by BBN is higher than the contribution from luminous baryons
so that there also exists a dark matter problem for baryons - a lot of
baryonic matter has to be hidden.

\subsection{The Milky Way disk}
The distinguishing feature between baryonic and non-baryonic matter is
that the former, due to its coupling to the electromagnetic field,
is able to emit light, i.e. radiate. As a consequence of this, energy is
dissipated which is why baronic matter in galaxies usually
is concentrated
to a bulge and/or bar near the center, and for spiral galaxies also
in the form of a thin disk. Non-baryonic dark matter, however,
most plausably consists of electrically neutral, non-interacting particles
and is therefore unable to condense by dissipation. On the other hand,
dark matter structure can evolve in several ways under the influence
of gravity. Besides the hierarchical clustering seen in simulations
of dark matter structure formation, there will also be an interplay
between the baryonic and non-baryonic components, e.g., infall and
accretion of dark matter and gas onto galaxies and perhaps also on
smaller substructures. It may therefore be interesting to estimate
the amount of dark matter also on the smaller scales represented, e.g.,
by the galactic disks.

For the Milky Way, Oort \cite{oort} was the first to give
 estimates based on the motions
of stars perpendicular to the disk. This analysis
 seemed to necessitate dark matter
in the solar neighbourhood in the disk. After some controversy, this
suggestion was however convincingly refuted by Kuijken and Gilmore
\cite{kuijken} who devised a method which used the full three-dimensional
distribution of velocities and positions of the tracer stars. They
showed that the dynamical effects on the tracers can be entirely accounted
for by visible material, leaving little room for disk dark matter (see
\cite{ashman} for an extensive review of the evidence for dark matter
on various scales in galaxies).

\subsection{Hidden baryons - gas}

Apart from the relatively large amounts of hot X-ray emitting
gas in clusters, the main
baryonic component of the universe may be
 diffusively distributed gas inbetween galaxies
and clusters. This is very difficult to detect at low
redshifts using presently
available methods. However, an analysis of the absorption of light
from distant quasars due to intervening gas shows that the amount of
hydrogen and helium gas corresponds well to the nucleosynthesis
prediction \cite{hogan}.
\subsection{Hidden baryons - MACHOs}

It is not excluded that a large amount of baryonic mass may be hidden
in galactic halos in the form of sub-solar mass objects, MACHOS
\cite{alcock}.
For such objects, the clever technique of microlensing has
been used as a tool. The idea is to monitor
1 to 10 million stars in a satellite galaxy, e.g., the Large
Magellanic Cloud (LMC) or Small Magellanic Cloud (SMC).
The intensity of one of these
background stars
will rise in a typical, time-symmetric and achromatic fashion
during a few days, weeks,
or months (depending on the mass and transverse velocity
of the intervening object)
if an object
such as a non-luminous star passes the line-of-sight to the star
\cite{paczynski}, if such stars make up a sizeable fraction
of the Galactic halo.
Indeed, such microlensing events were reported by two collaborations
\cite{macho,eros} soon after the observational programmes
started.

However, given the  optical depth
for microlensing observed towards the LMC \cite{eros,macho2},
the most likely fraction of
the halo mass given by MACHOs is not larger than around 20\%.  In
fact, it could be much smaller if debris from tidal stripping of the LMC itself
or other dwarf satellites happens to lie in the line-of-sight, as
indicated by some observations \cite{zaritsky}, or if LMC is more elongated
along the line-of-sight than previously thought \cite{spiro}.

A problem with the MACHO hypothesis, even if the events are due to
a halo population, is to understand what the lensing objects could be
\cite{freese,carr1}.
As even low-mass stars ($\sim 0.1M_\odot$)
radiate at some level, one may use the
very stringent limits from photometric surveys \cite{uson} of
Abell custers to rule
out a halo contribution of more than a percent of those objects.
Also, if stellar remnants are cosmologically important as dark matter
objects, their progenitors would have generated a diffuse infrared flux
throughout the universe, which would cause absorption of TeV
gamma rays. As such gamma rays have been observed from objects at
redshifts above 0.03, a stellar remnant population of the
required abundance seems very unlikely \cite{freese2}.

Since
the EROS microlensing experiment \cite{eros} rules out the region of
less massive stars (down to $10^{-5}$ solar masses), it is
difficult to fit conventional star-like objects into the hypothesis
of a MACHO halo, regardless of the very unusual mass function that
would be needed. In fact, evidence against a non-standard mass function
for stars is mounting, since recent measurements of the distribution
of stellar masses in a wide variety of environments indicate a universal
function, and one which supports the canonical view that low-mass objects
cannot be numerous enough to be dynamically important in galaxy halos
\cite{sofia}.

Thus, despite early optimism of at least a partial
solution of the dark matter problem when the first
events were reported using this elegant
microlensing technique, it may seem now that these
experiments are of more interest to stellar population studies, and
for determining the distribution of stars in the Milky Way and in the
Magellanic Clouds.
\subsection{Exotic baryonic matter}

Even though there is not much at present which points to a
major component of dark matter being of baryonic form,
one cannot entirely rule out some exotic scenarios.
Since many of these invoke ``conventional'' matter,
such as difficult to detect (and exclude)
cold molecular clouds or very low mass stars with an extremely
fine-tuned mass function, a major yet unsolved
problem for these models is to circumvent the nucleosynthesis
bound (or to adopt the unlikely hypothesis that we live
in a very low density universe).

A form of baryonic dark matter which could avoid the BBN
bound is primordial black holes \cite{ivanov}. If such objects formed
before nucleosynthesis (e.g., at the quark-hadron phase
transition) they would not have a noticeable effect
on the light element abundances. There are, however, many
problems with such a scenario. Recent thinking in particle
physics puts the strongly first order transition needed
at low probability. Also, the mass spectrum of primordial
black holes must be peaked at those particular masses
where observational constraints happen to be the weakest.

\section{Distribution of dark matter}\label{sec:distribution}
\subsection{Galactic halos}\label{subs:halos}
On galactic scales and smaller, the classical tests of the mass
distribution provided by rotation curves continue to be refined. A
compilation of almost 1000 rotation curves led to the
conclusion that dark matter indeed is present in large amounts \cite{persic}.
The problem of how
dark matter is distributed in halos of galaxies and galaxy clusters
is an important one for the purpose of determining strategies for the
detection of the various candidates, as we will see. Unfortunately,
the available data on the structure of the Milky Way do not constrain
the dark matter halo density profile very well \cite{dehmelt}.

For a spiral galaxy which has a spherically symmetric overall
mass distribution,
Newton's laws of gravity give for the rotation velocity of a tracer
star, or neutral hydrogen, at distance $r$ from the centre
\beq
{v_{\rm rot}^2\over r}= {G_NM(r)\over r^2},
\eeq
where $v_{\rm rot}$ is the rotation velocity and $M(r)$ is the
total mass of the galaxy interior to $r$.
This gives
\beq
M(r)={v_{\rm rot}^2r\over G_N},
\eeq
so that a constant rotation  velocity, which
is usually observed for spiral galaxies over a large range
of $r$  implies a halo mass which grows
linearly with $r$. Realistically, this growth can not extend
arbitrarily far. For mass distributions obtained from
numerical simulations of structure formation with cold dark matter,
the growth of mass eventually becomes only logarithmic until a halo
starts to overlap with one of a nearby galaxy.

A phenomenological form of the dark matter halo mass density distribution
with free parameters which can be chosen to reproduce most
measured rotation curves is given by

\beq
\rho(r)\propto {\rho_c\over (r/a)^\gamma
\left[1+(r/a)^\alpha\right]^{(\beta-\gamma)/\alpha}},\label{eq:halofamily}
\eeq
where $a$ is a dimensionful parameter related to the core radius
of the halo.
Among the many shapes contained in this family of density curves can
be mentioned the isothermal profile
with a core (model $S_p$ in the following), $(\alpha,\beta,\gamma)=
(2,2,0)$; the Navarro-Frenk-White (NFW) model $(\alpha,\beta,\gamma)=
(1,3,1)$, and the mildly singular models found in \cite{kravtsov},
$(\alpha,\beta,\gamma)=(2,3,0.2)$ (model $K_a$),
$(\alpha,\beta,\gamma)=(2,3,0.4)$ (model $K_b$).
All of these are capable of reproducing
the observed rotation curves of most galaxies over a large range of
radii, but have quite different behaviour at very small or very
large radii.

\begin{figure}[!htb]\begin{center}
\epsfig{file=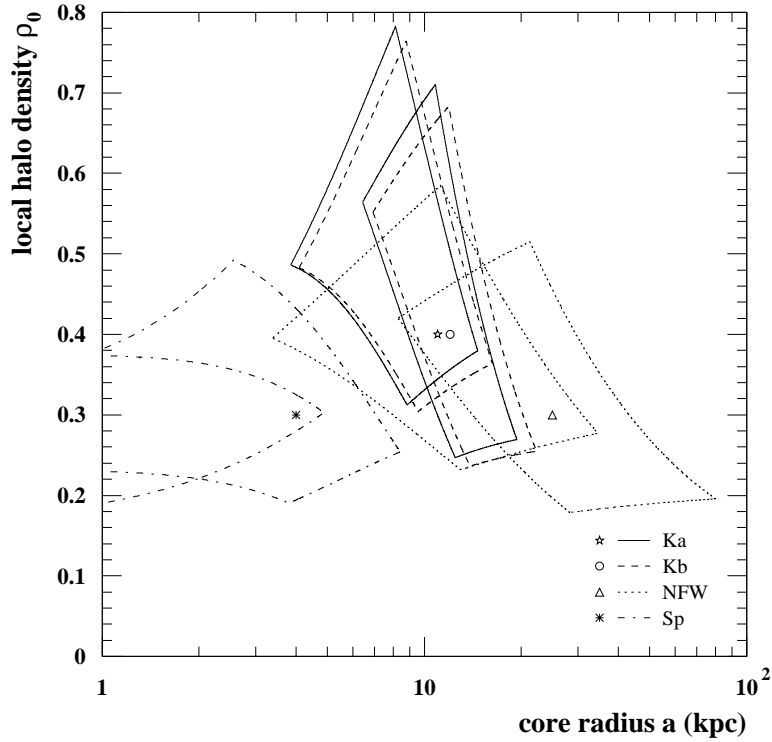,width=12cm}
\end{center}
\caption{Allowed values of the parameters $\rho_0$, the local halo density,
and $a$, the core radius, for some commonly used halo profiles
discussed in the text.
The allowed regions depend
on the galactocentric distance
of the solar system $R_0$; plotted are regions corresponding to
$R_0 = 8.5\, \rm{kpc}$ which extend to lower values of $a$ and
to $R_0 = 7.1\, \rm{kpc}$ which allow higher values of $a$. The markers indicate the halo profiles which are considered in Section~\protect\ref{subs:lines}.
Figure prepared by P. Ullio, for more details, see \protect\cite{BUB}.}
\label{fig:rholocal}
\end{figure}

If the Galactic halo consists of non-baryonic dark matter
particles, the local value $\rho_0$
of the dark matter density at our galactocentric distance $R_0$
 is of interest to various detection experiments.
Both
of these quantities are presently uncertain, in particular $\rho_0$
which also depends on the shape of the density profile.
In Fig.\,~\ref{fig:rholocal}, from \cite{BUB}, it is shown how $\rho_0$
is changing as a function of the ``core radius'' $a$ for some of
the models contained in the family given in Eq.~(\ref{eq:halofamily})
which can
describe the gross features of our Galaxy. As can
be seen, the variation is large but a range of
\beq
\rho_0\sim 0.2 - 0.5  \ {\rm GeV/cm}^3\label{eq:rholocal}
\eeq
covers most of the possibilities. In this analysis, a spherical
halo has been assumed. If a rotating and/or flattened halo is considered,
the range of reasonable values of the local halo density may be
somewhat larger \cite{gyuk,kink}.

Usually one takes the local galactic velocity distribution of
the dark matter particles to be a
truncated gaussian, which in the Earth frame moving at speed $v_O$
relative to the galactic halo means
\beq
  f(v) = {1\over {\cal N}_{\rm cut}} { v^2 \over u v_O \sigma} \left\{
  \exp\left[{-{(u-v_O)^2\over2\sigma^2}}\right] -
  \exp\left[{-{\min(u+v_O,v_{\rm cut})^2\over2\sigma^2}}\right]
  \right\}
\eeq
for $ v_{\rm esc} < v < \sqrt{v_{\rm esc}^2 + (v_O + v_{\rm cut} ) ^2
} $ and zero otherwise, with $ u = \sqrt{v^2 + v_{\rm esc}^2} $ and
\beq
  {\cal N}_{\rm cut} =
  {v_{\rm cut}\over\sigma} \exp\left( {-{v_{\rm cut}^2\over2\sigma^2}} \right)
  -
  \sqrt{\pi\over2} {\rm erf} \left( {v_{\rm cut}\over\sqrt{2}\sigma} \right) .
\label{eq:veldist}
\eeq
Typical values are for the halo line-of-sight velocity dispersion
$\sigma = $150 km/s, the galactic escape speed $ v_{\rm cut} = $ 600
km/s, the relative Earth-halo speed $ v_O = $ 230 km/s (a yearly
average) and the Earth escape speed $ v_{\rm esc} = $ 11.9 km/s.
The small variation of the relative Earth-halo speed with season
will cause a possible annual modulation in the detection
rate in Earth-based detectors for dark matter particles \cite{drukier}
as discussed further in
Section~\ref{subs:direct}.

\subsection{Dwarf and low surface brightness galaxies}

A very interesting class of
objects is provided by low surface brightness galaxies and
dwarf spiral and irregular
galaxies, which seem to be completely dominated by dark matter
\cite{gilmore}.
A couple of these have unusual rotation curves which
could perhaps be interpreted as being due to a combination of MACHOs
and nonbaryonic dark matter \cite{burkert_silk}.

\begin{figure}[!htb]\begin{center}
\epsfig{file=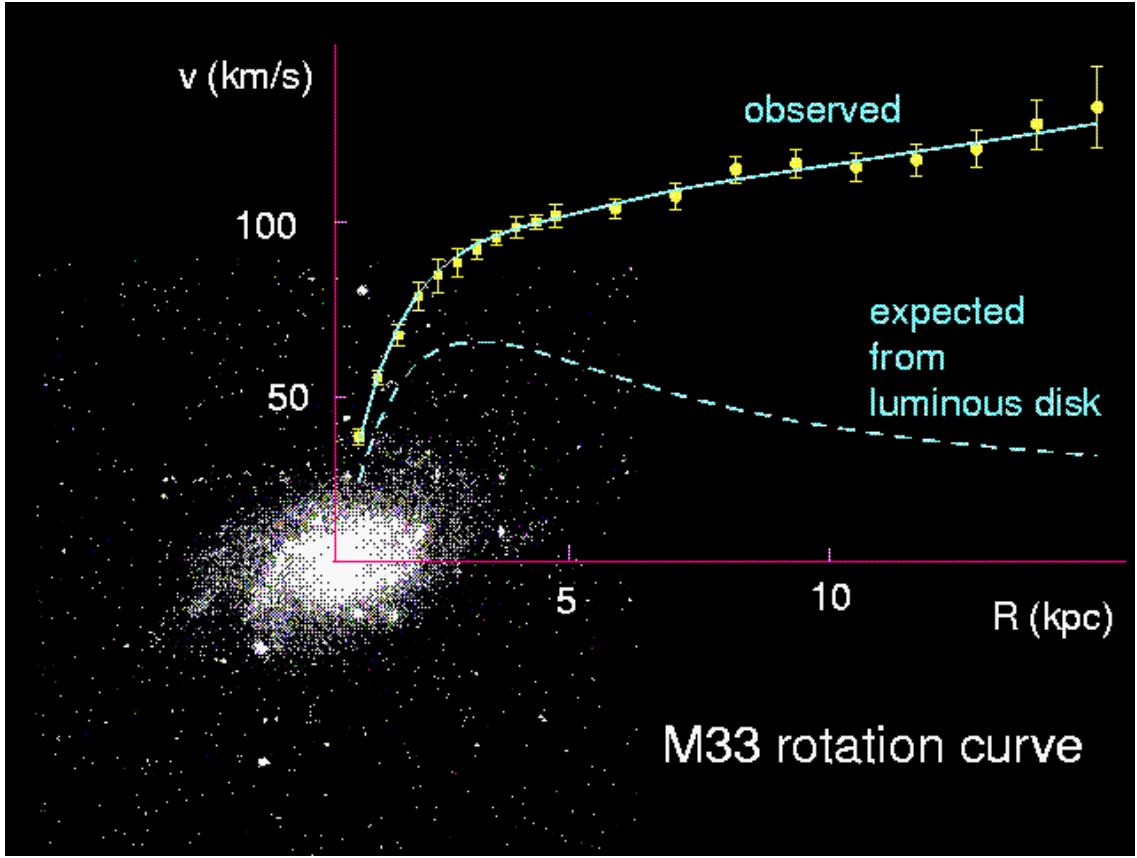,width=15cm}
\end{center}
\caption{Observed H I rotation curve of the nearby dwarf spiral galaxy M33
(adapted from \protect\cite{corbelli}), superimposed on an optical image \protect\cite{ned}.
The dashed line shows the estimated  contribution to the rotation curve from
the luminous stellar disk \protect\cite{corbelli}. There is also a smaller
contribution from gas (not shown).
}
\label{fig:m33}
\end{figure}

Dwarf spirals are interesting,  since they have rotation curves
which are clearly rising well beyond the scale length of the luminous disk.
In the local group, M33 at a distance of only around 0.8 Mpc gives a good
opportunity to study the rotation curve since it has HI gas which can
be traced out to large distances from the center. This is shown in
Fig.~\ref{fig:m33}, where the rotation curve measurements extracted
from \cite{corbelli} are shown superimposed on an optical image of
the galaxy. Also shown is the contribution to the rotation curve
from the luminous disk computed in \cite{corbelli}. (There is
at large radii also a small contribution from the gas mass, not shown
in the Figure.) As can be seen, the rotation curve is rising well beyond
the point where Newtonian dynamics based on only the luminous mass
would predict a decline. Since the curve continues to rise at the
last measured points, only a lower limit to the mass of the dark halo
can be given, $M\gsim 5\cdot 10^{10}\ M_\odot$, more than 10 times the
mass in stars and gas. It is noted in \cite{corbelli} that an NFW profile
in fact fits the rotation curve quite well but the central  concentration is
lower than that predicted by an $\Omega_M = 1$ universe, perhaps indicating
again the need to decrease the matter density (while allowing
a cosmological constant to give a flat large-scale geometry).

\section{Models for non-baryonic dark matter}\label{sec:models}
Given that the total mass density of the universe seems to be higher
than what is allowed by big bang nucleosynthesis for baryons alone,
an important task of
cosmology and particle physics is to produce viable non-baryonic
candidates and to indicate how the various scenarios can be tested
observationally.

\subsection{Changing the law of gravity?}

 It has
 turned out to be very difficult to modify gravity on the various length
 scales where the dark matter problem resides,
but phenomenological attempts have been made to at
least explain flat galaxy rotation curves by introducing
violations of Newton's laws (and of general relativity) \cite{milgrom}.
Until a satisfactory alternative theory to general relativity has
been found it is difficult to further comment on this option.
Besides the remarkable success of the ``standard'' theory
in accounting for perihelion motion, redshifts, gravitational lensing and
binary pulsar dynamics, the overall consistency of the standard
cosmology it provides the basis for, also on the largest scales,
 is remarkable. An example is the concordance of
the mass estimates of galaxy clusters based on galaxy velocity
dispersions, gravitational lensing, microwave background distorsions
and X-ray emission from hot intracluster
gas. At present, there
does not seem to exist a plausible alternative theory that can
match this impressive list of successes.

In principle, there are modifications to Newtonian gravity
if there exists a non-zero cosmological constant, since
the energy equation for a test particle of mass $m$
at a distance $R$ from a homogeneous sphere  of mass $M$
gets an additional term proportional
to $\Lambda$,
\beq
E={1\over 2}m\dot R^2-{G_N Mm\over R}-{\Lambda\over 6}mR^2,
\eeq
(see \cite{BGbook}) showing the attractive nature of the extra force
for $\Lambda<0$. However, this additional term is some four orders
of magnitude too small to have measurable effects
in galactic systems, given the current observational estimates of
$\Lambda$ \cite{cosm}. In addition, the observationally favoured value of
$\Lambda$ is positive and thus causes repulsion instead of attraction.

\subsection{Particle Dark Matter}\label{subs:particledm}
We have seen that many independent observations point to
the existence of non-baryonic dark matter in the universe. Even in
the absence of observations, we noticed in Sec.\,~\ref{sec:relic}
that it may not be unexpected that massive, electrically neutral,
weakly interacting and
long-lived particles make up a substantial fraction of the
average cosmic mass density.
If such a massive particle species has roughly the same
type of gauge couplings as the known quarks and leptons, it must
have been produced in large abundances in the earliest universe
when the thermal energies were high enough to produce it in
collisions between ordinary particles.

As the universe expanded, the temperature decreased and eventually
the production was cut off because of the lack of sufficient energy
of the colliding particles in the primordial plasma. Also,
the probability that particles of this new type would collide
with each other and annihilate decreased rapidly as their
number density was diluted by the expansion. This generic
mechanism was investigated first for neutrinos
\cite{hut,leeweinberg,vysotsky,gunn}
but is generally valid, also for heavier particles as discussed
in Section~\ref{sec:relic}.

There are slight modifications needed for various
types of possible dark matter candidates. For example, if the
particle is different from the antiparticle, it may be that there
exists an asymmetry which can make the relic number density higher
than if there would be a perfect symmetry. This may allow for
a relic density which is higher than the estimate in
Eq.~(\ref{eq:relicdensity})
even if the annihilation cross section is large. In fact, such an
asymmetry must have existed for the baryons, because otherwise
baryons would have been almost completely annihilated by antibaryons
due to the large annihilation cross section caused by the strong interaction,
making a very different universe from that observed.
From BBN considerations,
one finds that the baryon-antibaryon asymmetry must have been of
the order of one part in $10^9$. The origin of this CP violating
and baryon number violating asymmetry is still unknown, and probably
one needs to go beyond the standard model of particle physics to explain
it.

If the particle is identical to the antiparticle,
this uncertainty of the amount of asymmetry does not appear.
The electrically neutral spin-1 particles $\gamma$ and $Z^0$
are examples of such self-charge conjugate particles.
It is possible also for a neutral spin-1/2 particle to be
its own antiparticle, a so-called Majorana fermion. This is
the generic case for supersymmetric particles to be discussed
in Section~\ref{sec:susy}.

For Majorana and other self-conjugate particles, and for particles where
the asymmetry is zero, the calculation of the relic density proceeds
as in Section \ref{sec:relic}, and the relic density can usually
be estimated from Eq.\,(\ref{eq:relicdensity}).

Assuming a thermal production process in the early universe, there is
an upper limit to the mass of a stable relic  particle \cite{unitarity}.
This comes about beacuse unitarity precludes the annihilation cross section
of particles of mass $M$, spin $J$
and relative velocity (in the centre of mass frame)
$v_{\rm rel}$ from being larger than $4\pi(2J+1)/(M^2v_{\rm rel}^2)$.
Using the estimated $v_{\rm rel}$ at freeze-out, it is found that
$M$ cannot exceed around 340 TeV. The most favoured heavy dark matter
candidate, the lightest supersymmetric particle, always has a mass much
below this limit in the minimal models. There may be a possibility
to evade the unitarity bound and accept even extremely heavy particles
as dark matter candidates if, for instance, they are not absolutely
stable (so that the formula in Eq.~(\ref{eq:relicdensity}) does not
apply), or if the production mechanism is non-thermal \cite{wimpzilla,crypton}.

\subsection{Particles with only gravitational interactions }
Although there are particle physics motivated dark matter candidates
which have non-negligible couplings to ordinary matter, and which
therefore are in principle detectable through other interactions
than gravity, there is always the possibility
that the dark matter interacts only gravitationally, or extremely weakly,
with ordinary matter. In fact, this possibility may arise
in different string theory-inspired scenarios \cite{dolgov2,nina}.
The possibility to confirm or experimentally rule out such presently very
speculative models is uncertain to say the least, and we will not
discuss them further.

\subsection{Massive neutrinos}
Of the many candidates for non-baryonic dark matter proposed,
neutrinos are often said to  have the undisputed virtue of being known to exist.
Actually, this is  a statement which needs some qualification because
neutrinos can only be dark matter candidates if they are massive. For
this to be true, both left-handed and right-handed neutrino states
are needed, and the latter are not known to exist (in the minimal
Standard Model of particle physics the right-handed neutrino is simply
absent). In principle, one can construct a mass term from only the
left chirality neutrino field, but this gives a Majorana type mass
which violates lepton number by two units, and in the Standard Model
$B-L$ is exactly conserved. Also, one would have to introduce an
isotriplet Higgs field in addition to the isodoublet present in the Standard
Model.

Non-zero neutrino masses, if established, would thus be an indication
of physics beyond the Standard Model. Since there exists a number of
indications that the Standard Model cannot be the final theory, it
would not be a big surprise if neutrinos are massive. As the direct
experimental limits on neutrino mass show \cite{pdg},
\begin{center}
\beqa
m_{\nu_{e}}&<&15\ {\rm eV}\nonumber\\
m_{\nu_{\mu}}&<&0.19\ {\rm MeV}\nonumber\\
m_{\nu_{\tau}}&<&18.2\ {\rm MeV}\nonumber\\
\eeqa
\end{center}
the neutrino masses have to be much smaller than the corresponding
quark and charged-lepton masses.
An intriguing explanation of this
fact could be given by the so-called see-saw mechanism,
where a right-handed Majorana mass $M$, at a large scale $\propto
M_{GUT}\sim 10^{15\pm 2}$ GeV  modifies through mixing the usual Dirac-type mass
$m_{D}$ of the lightest state  to $m_{D}^2/M\ll m_{D}$. In the
simplest versions of this
scheme, the neutrino masses would scale as the square of the
corresponding charged-lepton masses. There are variants (e.g., in
models of loop-induced neutrino masses) where neutrino masses
are instead linearly related to the charged-lepton masses.

Indeed, there exist several indications that neutrinos are not
massless. Although the direct kinematical measurements of
neutrino masses have given values consistent with zero,  evidence
from neutrino oscillation experiments is mounting that neutrinos
oscillate in flavour and hence must posses non-zero masses. To give a
cosmologically interesting contribution to $\Omega$, a relatively
narrow range $m_{\nu}\sim 1-50$ eV is required
(see Eq.\,~(\ref{eq:relicneutrino})). A neutrino heavier
than that would overclose the universe unless $m_{\nu}> 3$ GeV, when it would be
non-relativistic at freeze-out with a small enough relic abundance to
act as Cold Dark Matter.
This is ruled out  for Dirac neutrinos by  accelerator and direct detection data
up to the TeV range. At the other mass end,
a neutrino lighter than 1 eV would only give a small and dynamically not very
important contribution to $\Omega$.

Of the various experimental indications of neutrino oscillations,
only the LSND results \cite{caldwell} seem to be in the cosmologically
interesting range, with $\Delta m^2\sim 1 - 6$ eV$^2$. These results, however,
need independent
confirmation from other experiments. In fact, large portions of the
region of mass differences and mixing angles indicated by LSND
have been excluded by the KARMEN experiment \cite{karmen}. A definitive answer
will probably have to await new experiments such as  BooNE
at Fermilab \cite{boone}.

The solar neutrino problem,
which in view of new helioseismological data
does not seem to be solvable by changing  the standard
astrophysical solar model
\cite{dalsgaard},
and thus presents rather compelling evidence for oscillations, indicates
solutions with very small $\Delta m^2$.
 This would imply
small absolute values of neutrino masses unless there exists a mass
degeneracy of unknown origin between neutrinos. Such a degeneracy
is, however, easily destroyed by higher order quantum corrections, and
therefore seems contrived \cite{ellisdeg}.

Likewise, the
atmospheric neutrino anomaly, recently confirmed by Super-Kamiokande
data, has a preferred solution with a $\Delta m^2$ of only a few
times $10^{-3}$ eV$^2$. Seen already in the smaller Kamiokande
detector \cite{kamosc}, as well as in IMB \cite{imb} and recently also
confirmed by Soudan-2 \cite{goodman} and MACRO \cite{macro},
what is observed is a deficit
in the ``ratio of ratios'', $r\equiv (\nu_{\mu}/\nu_{e})_{data}
/ (\nu_{\mu}/\nu_{e})_{MC} \sim 0.6$. The interpretation in terms of
neutrino oscillations is particularly  compelling with the Super-Kamiokande
data where a zenith-angle dependence of the ratio is indicated
\cite{totsuka} with higher significance than in the other experiments.
Thus it is very likely that neutrinos are
indeed massive, but the mass is too small to be very significant for
cosmology (although, as noted in the Introduction, neutrinos most
probably contribute
as much to the energy density in the universe as the visible stars).
However, even if the largest neutrino mass
is  of the order of only a few tenths of an eV, as indicated by the
atmospheric neutrino anomaly, the effects on structure formation could
still be large enough to be detected in the far future by combining
data from large galaxy surveys similar to the Sloan Digital Sky
Survey, with precision measurements of the cosmic microwave
background from the future Planck satellite \cite{huetal}.

There is a fundamental objection to having massive
but light neutrinos as
the dominant
constituent of
dark matter on all scales where it is observationally needed. This
has to do with the fact that neutrinos are spin-$1/2$ particles
obeying the Pauli exclusion principle. To make up the dark matter
in dwarf galaxies (which are observed to be completely dominated
by the dark matter component, see Fig.\,\ref{fig:m33}), neutrinos would have to be stacked
together so tightly in phase-space that it is impossible to evade
the Pauli principle. Quantitatively, Tremaine and Gunn found \cite{tg}
that to explain the dark matter of a dwarf galaxy of velocity
dispersion $\sigma$ (usually of order 100 km/s) and core radius $r_{c}$
(typically 1 kpc), the neutrino mass has to
fulfil
\beq
m_{\nu}\geq 120\ {\rm eV}\left({100\ {\rm km/s}\over\sigma}\right)^{{1\over
4}}\left({1\ {\rm kpc}\over r_{c}}\right).\label{eq:gt}
\eeq

This high value is, however, not consistent with the requirement
$\Omega_{\nu}h^2\leq 1$, which according
to Eq.~(\ref{eq:relicneutrino}) requires $\sum_{i}m_{\nu_{i}}< 93$ eV.
The more desirable values $\Omega_\nu\sim 0.25$, $h\sim 0.65$ give in fact
$m_\nu\sim 10$ eV, which violates Eq.~(\ref{eq:gt}) by an even larger amount.
One way out of this particular problem would be if a 10 eV scale neutrino
exists and is unstable on the comological time scale. This, however,
requires exotic decay modes, since weak interaction mediated decays
occur on a much longer timescale. The problem with all models for the
dark matter which rely on decaying relic particles is that
a considerable amount of fine tuning of the product of relic density
and decay time is needed to obtain sensible values of $\Omega_M$.

We mentioned
that the upper bound in Eq.~(\ref{eq:relicneutrino})
on the neutrino mass is only applicable for ``standard'',
very light neutrinos. As a consequence of the behaviour of the
freeze-out abundance and cross section as a function of mass, there
is another mass range around 3 GeV where the relic density would
be close to critical. Today this window is ruled
out by the precision measurements at the LEP accelerator at
CERN where the
number of neutrino species which couple in the usual way to the $Z$ boson
has been determined to be three.
The pre-LEP papers which worked out the dark matter phenomenology
of such massive neutrinos (e.g., \cite{hut,leeweinberg,vysotsky,gunn})
were important, however, since they showed that a weakly interacting,
massive particle
(``WIMP'') could serve as cold dark matter with the required relic density.

To conclude this section about neutrinos,
it seems that
it is very plausible that they
make up some of the dark matter in the universe (given the
experimental results on neutrino oscillations), but most of the dark matter
is of some other form. Particle physics offers several promising
candidates for this.

\subsection{Axions}\label{subs:axions}

In particle physics, the combined action of charge conjugation (C) and
parity (P) is not an exact symmetry. For instance, higher-order weak
interactions involving quarks from all three generations are believed
to cause the experimentally observed small CP violation in the
neutral K meson system. With the advent of quantum chromo dynamics (QCD)
as the fundamental gauge theory for the strong interaction, it
was found that non-perturbative effects should induce a much larger
CP violation in the strong sector. However, the absence, e.g., of an
electric dipole moment of the neutron puts severe upper limits on
such a  strong CP-violating parameter.  The idea of Peccei and Quinn
was to make the CP violating phase dynamical \cite{PQ} by
introducing a global
symmetry, $U(1)_{PQ}$, which is
spontaneously broken. The Goldstone
boson of this broken global symmetry is the axion, which however gets a
non-zero mass from the QCD anomaly, which can be interpreted as a
mixing of the axion field with the $\pi$ and $\eta$ mesons
\cite{weinax,wilcax}.

The earliest attemps, using only the standard model particles
but with an enlarged Higgs sector, were soon ruled out experimentally
and the ``invisible axion'' was invented \cite{DFSZ,KSVZ} with a very high mass
scale of symmetry breaking and with very massive fermions
carrying PQ charge. This means that  only a feeble strong or electromagnetic
interaction leaks  out to the visible sector through triangle loop
diagrams.

The
phenomenology of the axion is determined, up to numerical factors, by
one number only - the scale $f_{a}$ of symmetry breaking. In
particular, the mass is given by
\beq
m_{a}=0.62\ {\rm eV}\left({10^7\  {\rm GeV}\over f_{a}}\right),
\eeq
and the experimentally important coupling to two photons is due to
the effective Lagrangian term
\beq
{\cal L}_{a\gamma\gamma}
=\left({\alpha_{\rm em}\over 2\pi f_{a}}\right)\kappa{\bf E\cdot B}a,\label{eq:axcoup}
\eeq
where ${\bf E}$ is the electric field, ${\bf B}$ is the magnetic field and
$\kappa$ is a model-dependent parameter of order unity.

\begin{figure}[!htb]\begin{center}
\epsfig{file=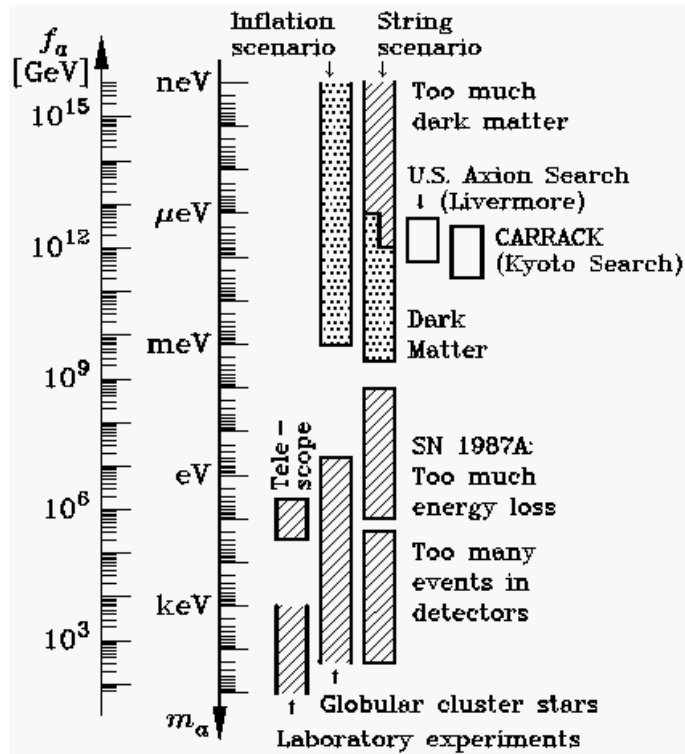,width=9cm}
\end{center}
\caption{\label{fig:axion} Astrophysical and cosmological exclusion
regions (hatched) for the axion mass $m_a$ or equivalently, the
Peccei-Quinn scale $f_a$. An ``open end'' of an exclusion bar means
that it represents a rough estimate.
The dotted ``inclusion regions'' indicate where axions could plausibly
be the cosmic dark matter.  Most of the allowed range in the inflation
scenario requires fine-tuned initial conditions.  In the string
scenario the plausible dark-matter range is controversial as indicated
by the step in the low-mass end of the ``inclusion bar.'' Figure kindly
provided by G. Raffelt; for more details see \protect\cite{raffelt} and
\protect\cite{georgreview}.}
\end{figure}

The axion, constrained by laboratory searches, stellar cooling
 and the dynamics of supernova 1987A to be very light, $m_{a}< 0.01$
 eV \cite{raffelt}, couples so weakly to other matter
\cite{invisible} that it never was in thermal equilibrium in the early
universe and it would
 behave today as Cold Dark Matter. The window where axions are viable
 DM candidates is progressively getting smaller, but still there is an
 acceptable range between around $10^{-5}$ and $10^{-2}$ eV where they
 pass all observational constraints and would not overclose the universe,
see Fig.~\ref{fig:axion} taken from \cite{georgreview}.
There is a considerable uncertainty in the relation
between mass and relic density,
depending on the several possible sources of axion production such as vacuum
misalignment, emission from cosmic strings etc.

The coupling in Eq.~(\ref{eq:axcoup}) implies that resonant conversion
between a galactic axion
and an electric photon mode may take place in the presence of a strong
magnetic field -
not even the ``invisible axion'' may be undetectable \cite{sikivie_ax},
since the number density of these light particles in the Galaxy has to be
enormous
if axions are to make up the dark matter.

 Fortunately, there are now  two experiments \cite{llnl,kyoto}
 which have the experimental sensitivity of probing much of the interesting
 window within the next few years.

 Axions share with massive neutrinos and the supersymmetric candidates to be
 discussed next the attractive feature of having other,
 particle-physics motivated, reasons to exist besides giving a
 possible explanation of dark matter. Of course there are other
 proposed candidates which, although not yet generally accepted, could
 finally turn out to give the correct explanation.

\section{Weakly interacting massive particles - supersymmetric
particles}\label{sec:susy}
One of the prime
candidates for the non-baryonic component is provided by the lightest
supersymmetric particle, plausibly the lightest neutralino $\chi$, to
be described further in Section~\ref{sec:mssm}.

Supersymmetry seems to be a necessary ingredient in superstring theory (which is now
commonly seen as one aspect of a grander theory,
M-theory) which unites all the
fundamental forces of nature, including gravity. In most versions of
the low-energy theory  there is a conserved multiplicative quantum
number, R-parity:
\beq
R=\left(-1\right)^{3(B-L)+2S},
\eeq
where $B$ is the baryon number, $L$ the lepton
number and $S$ the spin of the particle. This implies that
$R=+1$ for ordinary particles
and $R=-1$ for supersymmetric particles. This means that supersymmetric
particles can only be created or annihilated in pairs in reactions
of ordinary particles. It also means that a single supersymmetric particle
can only decay into final states containing an odd number of
supersymmetric particles. In particular, this
makes the lightest supersymmetric particle
stable, since there is no kinematically allowed state with negative
R-parity which it can decay to.

Thus, pair-produced neutralinos in the early universe which
left thermal equilibrium as the universe kept expanding should have a non-zero
relic abundance today. If the scale of supersymmetry breaking is related to
that of electroweak breaking, Eq.\,~(\ref{eq:relicdensity})
shows that $\Omega_{\chi}$ will be
of the right order
of magnitude to explain the non-baryonic dark matter.
It would indeed appear as an economic
solution if  two of the  most outstanding problems in fundamental
science, that of dark matter and that of the unification of the basic
forces, would have a common element of solution - supersymmetry.

The idea that supersymmetric particles could be good  dark matter
candidates became attractive when it was realised that breaking
of supersymmetry could be related to the electroweak scale, and
that, e.g., the supersymmetric partner of the photon (the photino)
would couple to fermions with electroweak strength \cite{fayet}.
Then
most of the phenomenology would be similar to the (failed) attemps
to have multi-GeV neutrinos as dark matter. After some early
work along these lines \cite{cabibbo,primack,weinberg2,goldberg,krauss},
the first
more complete discussion of the various possible supersymmetric
candidates was provided in \cite{ellis0}, where in particular the
lightest neutralino was identified as perhaps the most promising one.

Supersymmetric  dark matter should be seen
 at one particular realization of a generic WIMP
(weakly interacting massive particle). Here weakly
 interacting, electrically neutral massive (GeV to TeV range)
 particles are assumed to carry a conserved quantum number ($R$-parity
 in the case of supersymmetry) which
 suppresses or forbids the decay into lighter particles. Such
 particles should have been copiously produced in the early universe
 through their
 weak interactions with other forms of matter and radiation. As the
 universe expanded and cooled, the number density of the WIMPs
 successively became too low for the annihilation processes to keep
 up with the Hubble expansion rate. A relic population of WIMPs should
 thus exist, and it is very suggestive that the canonical weak
 interaction strength is, according to detailed calculations,
 just right to make the relic density fall in the
 required range to contribute substantially to $\Omega$.

 In addition, we saw in Section~\ref{sec:relic} that
WIMPs are generically found to decouple at a temperature
of roughly $m_{\rm WIMP}/20$, which means that they are
 non-relativistic already at decoupling and certainly behave as CDM
 by the time of matter dominance and structure formation, which
seems to be preferred observationally.

\subsection{Supersymmetric particles}\label{sec:mssm}

Let us now focus on  the lightest
supersymmetric particle, which if R-parity is conserved, should be stable.
In some early work, a decaying photino \cite{cabibbo}
or a gravitino \cite{primack} were considered, but for various reasons
\cite{ellis0} the most natural supersymmetric dark matter candidate  is
the lightest neutralino $\chi$. Thus it is a mixture
of the supersymmetric partners of the photon, the $Z$ and the two neutral
$CP$-even Higgs bosons present in the minimal extension of the
supersymmetric standard model (see, e.g., \cite{haberkane}). The
attractiveness
of this candidate, besides its particle physics virtues,
stems from the fact that it is electrically
neutral and thus neither absorbs nor emits light, and stable so that it
 can have survived since the big bang.
 Furthermore, it has gauge couplings
 and a mass
which for a large range of parameters in the supersymmetric sector
imply a relic density in the required range to explain
the observed $\Omega_M\sim 0.3$. As we will see, its couplings to ordinary
matter also means that its existence as dark matter in our galaxy's halo
may be experimentally tested.

These are the good properties of neutralinos as dark matter candidates. Less
attractive is the fact that virtually nothing is known about how
supersymmetry is broken, and therefore any given supersymmetric
model contains a large number of unknown parameters (of the order of 100).
Such a large parameter space is virtually impossible to explore by
present-day numerical methods, and therefore simplifying assumptions
are needed. Fortunately, most of the unknown parameters
such as CP violating phases influence the properites relevant
for cosmology, and for detection, very little. (In some specific cases, the
effects of CP violation may be non-negligible \cite{cpviol}.)

Usually, when scanning the large space
of a priori unknown parameters in supersymmetry, one thus makes
reasonable simplifying assumptions and accepts solutions
as cosmologically appropriate if they give a neutralino relic
density in the range
\beq
0.025 \lsim \Omega_\chi h^2 \lsim 1.\label{eq:bound}
\eeq
The lower limit comes from the desire to at least explain the dark
matter halos of galaxies, and the upper limit is a (probably too
conservative) upper limit of the observed matter density.

It should be noted, however, that the dark matter may have several
components, and that supersymmetric dark matter could exist even
if the lower bound in Eq.~(\ref{eq:bound}) is violated. Since there is
in general a crossing symmetry relating a large annihilation
cross section (and therefore small $\Omega_\chi$) to large scattering
rates in detectors (and annihilation rates in the Galactic halo),
such models giving a small relic density may in fact be
interesting from the experimental point of view. Unfortunately there
is, however, no simple prescription of how to go from the value of the
relic density to the mass density in our Galactic halo (since that
depends on the unknown formation history of the Galaxy).
A phenomenological recipe has been \cite{rescale} to rescale
Eq.\,~(\ref{eq:rholocal}) by a factor $\Omega_\chi h^2/0.025$ if
the computed relic density is smaller than the lower bound
in Eq.~(\ref{eq:bound}). This linear rescaling of the local dark
neutralino density implies for a given
neutralino mass a corresponding linear decrease of scattering
rates in detectors (and a quadratic decrease of annihilation rates
in the halo, since the probability for two dark matter particles to
annihilate is proportional to the square of the number density).
It can be argued that the range in  (\ref{eq:bound}) is too
generous. To narrow down the number of suitable supersymmetric models
we will in the examples below sometimes use the more easily
motivated range $0.1 < \Omega_\chi h^2 < 0.2$ (see
Section~\ref{sec:amount}).
for the supersymmetric models to be considered and not invoke rescaling.

Besides its interesting implications for cosmology, the motivation
from particle physics for supersymmetric particles
at the electroweak mass scale
has  become stronger due to
the apparent need for 100 GeV - 10 TeV scale supersymmetry to achieve
unification of the gauge couplings in view of LEP results
\cite{amaldi}. (For an extensive review of the literature on
supersymmetric dark
matter up to mid-1995, see Ref.\,~\cite{jkg}.)

Thanks to exciting developments in string theory \cite{strings},
supersymmetry has become an even more attractive feature to be
expected at the doorstep beyond the Standard Model. At a more
phenomenological level, supersymmetry gives an attractive solution to the
so-called hierarchy problem, which is to understand why the
electroweak scale at a few hundred GeV is so much smaller than the Planck scale
$\sim 10^{19}$ GeV despite the
fact that there is nothing in non-supersymmetric theories to
cancel the severe quadratic divergences of loop-induced mass terms.
In supersymmetric theories, the partners of differing spin would
exactly cancel those divergencies (if supersymmetry were unbroken).
Of course, supersymmetric models are not guaranteed to contain good
dark matter candidates. In particular, $R$-parity may not be
a conserved symmetry \cite{valle}
in which case there may not exist a long-lived
enough particle to make up the dark matter. In the simplest models, however,
and in particular in the minimal supersymmetric extension of the
ordinary Standard Model that we now discuss, $R$-parity is conserved
and the neutralino is a good dark matter candidate.

\subsubsection{MSSM: The minimal supersymmetric extension of the standard model}

\par
The minimal supersymmetric extension
of the standard model is defined by the particle content and
gauge couplings required by supersymmetry and a gauge-invariant
so-called  superpotential. Thus, to each particle degree of freedom in the
non-supersymmetric Standard Model, there appears a supersymmetric partner
with the same charge, colour etc, but with the spin differing by half
a unit. The only addition to this  doubling of the
 particle spectrum of the Standard Model concerns the Higgs sector. It
 turns out that the single scalar Higgs doublet is not enough to give
 masses to both the $u$- and $d$-like quarks and their
 superpartners (since supersymmetry forbids using both
a complex Higgs field and its complex conjugate at the same time, which
one does in the non-supersymmetric Standard Model).
Thus, two complex Higgs doublets have to be
 introduced. After the usual Higgs mechanism, three of these states
 disappear as the longitudinal components of the weak gauge bosons
 leaving five physical states: two neutral scalar Higgs particles $H_{1}$ and
 $H_{2}$ (where by convention $H_{2}$ is the lighter state), one
 neutral pseudoscalar state $A$, and two charged scalars $H^{\pm}$.
 The $Z$ boson mass gets a contribution from the vacuum expectation
 values (VEVs) of both of the doublets, but the way this division is done
 between the VEV $v_{1}$ of $H_{1}$ and $v_{2}$ of
$H_{2}$ is not fixed a priori.

 Electroweak symmetry breaking is thus caused by the neutral
 components of both $H_1$ and $H_2$
acquiring vacuum expectation values,
\begin{equation}
  \langle H^1_1\rangle = v_1 , \qquad \langle H^2_2\rangle = v_2,
\end{equation}
with $g^2(v_1^2+v_2^2) = 2 m_W^2$, with the further assumption that
vacuum expectation values of all other scalar fields (in particular,
squark and sleptons) vanish. This avoids color and/or charge breaking
vacua.
 The ratio of VEVs
 \beq
 \tan\beta\equiv {v_{2}\over v_{1}}
 \eeq
 always enters as a free parameter in the MSSM, although it seems
 unlikely to be outside the range between 1.1 and 45 \cite{jkg}.

 After supersymmetrization, the theory  also has to contain the
 supersymmetric partners of the spin-0 Higgs doublets. In particular,
 two Majorana fermion states, higgsinos,  appear as the supersymmetric
 partners of the electrically neutral parts of the
$H_{1}$ and $H_{2}$ doublets. These can mix
quantum mechanically with each other and
 with two other neutral
 Majorana states, the supersymmetric partners of the photon (the photino)
 and the $Z$ (the zino). When diagonalizing the mass matrix of these
 four neutral Majorana spinor fields (neutralinos), the lightest physical state
 becomes an excellent candidate for Cold Dark Matter.

 The non-minimal character of the Higgs sector may well be the first
 experimental hint at accelerators of supersymmetry. At tree level,
 the $H^ 0_{2}$ mass is smaller than $m_{Z}$, but radiative (loop)
corrections are important and shift this bound by a considerable amount.
However, even after allowing for such
 radiative corrections it can hardly be larger than around 130 GeV.
The successful operation of the CERN accelerator
LEP at centre of mass energies above 200 GeV without observing any supersymmetric
particles puts important constraints on the parameters of the MSSM.
Besides the limits on the Higgs masses (presently above 100 GeV for
the Standard Model Higgs), constraints on the chargino mass are
significant for many dark matter searches. It has proven to be
very difficult, however, to put very tight lower limits on the
mass of the lightest neutralino, because of the multitude of
couplings and decay modes of the next-to-lightest supersymmetric
particle. The lightest neutralino can in general only be
detected indirectly in accelerator experiments
through the missing energy and momentum it
would carry away from the interaction region.
As an example of current limits, or low values of $\tan\beta$, the present
 lower limit on the mass of the lightest neutralino
from the ALEPH collaboration is around 37 GeV
\cite{blondel}.

The upper limit of dark matter neutralino masses in the MSSM consistent
with Eq.~(\ref{eq:bound}) is
of the order of 7 TeV \cite{coann}. Above that mass, which is still
far from the unitarity bound of  340 TeV, the relic density becomes
larger than the upper limit in Eq.~(\ref{eq:bound}).
To get values for the lightest neutralino mass larger than a few
hundred GeV, however, some degree of ``finetuning'' is necessary
\cite{finetune}.
By making additional well-motivated but not mandatory
 restrictions on the parameter space, such as
in supergravity-inspired models, one gets in general masses
below 600 GeV \cite{ellisco} for the lightest neutralino.
\subsubsection{Supersymmetry Breaking}

Supersymmetry is a mathematically beautiful theory, and would give
rise to a very predictive scenario, if it were not broken in an
unknown way which unfortunately introduces a large number of unknown
parameters.

Breaking of supersymmetry has of course to be present
since no supersymmetric particle has as yet been
detected, and unbroken supersymmetry requires
particles and sparticles to have the same mass. This breaking can be
 achieved in the MSSM by a soft
supersymmetry-breaking potential which does not re-introduce large
 radiative mass-shifts (and which strongly indicates that the
 lightest supersymmetric particles should not be too much heavier than the 250
 GeV electroweak breaking scale). The origin of this effective
 low-energy  potential need not be specified,
 but it is natural to believe that it is induced through explicit
 breaking in a hidden sector of the theory at a high mass scale. The
supersymmetry breaking terms are then transmitted to the visible sector
 through gravitational interactions.

 Another possibility is that
 supersymmetry breaking is achieved through gauge interactions at
 relatively low energy in the hidden sector \cite{gauge}. This is then
 transferred to the visible sector through some messenger fields which
 transform non-trivially under the Standard Model gauge group.
 Although this scenario has some nice features, it does not  seem to give
 as natural a candidate for the dark matter as the ``canonical''
 scenario, which is the one we shall assume in most of the following.
 See, however, Ref.~\cite{gaugedm} for some possibilities of dark matter
 candidates in gauge-mediated models.

 Since one of the virtues of supersymmetry is that it resurrects the
 hope for grand unification of the gauge interactions at a common mass scale,
  a simplifying assumption based on this unification is often used
for the gaugino mass parameters,
\begin{equation}
  \begin{array}{l}
  M_1 = {5\over 3}\tan^2\theta_wM_2\esim 0.5 M_2, \\
  M_2 = { \alpha_{\rm em} \over \sin^2\theta_w \alpha_s } M_3 \esim 0.3 M_3,
  \end{array}
  \label{gauginounif}
\end{equation}
where $\theta_W$ is the weak mixing angle, $\sin^2\theta_W\approx 0.22$.

As mentioned, the one-loop effective potential for the Higgs fields
 has to be used used to obtain  realistic Higgs mass estimates.
 The minimization conditions of the potential allow one to
trade two of the Higgs potential parameters
for the $Z$ boson mass $m_Z^2 = {1\over2} (g^2+g'^2)
(v_1^2+v_2^2)$ (where $g=e/\sin\theta_W$, $g'=e/\cos\theta_W$) and the ratio of VEVs, $\tan\beta$.
The third parameter
can further be reexpressed in terms of the mass of one of the physical
Higgs bosons, for example $m_{A}$.

The neutralinos $ \tilde{\chi}^0_i$ are linear combination of the
neutral gauge bosons ${\tilde B}$, ${\tilde W_3}$ (or equivalently
$\tilde\gamma$, $\tilde Z$) and of the neutral
higgsinos ${\tilde H_1^0}$, ${\tilde H_2^0}$.  In this basis, their
mass matrix
\begin{eqnarray}
  {\cal M} =
  \left( \matrix{
  {M_1} & 0 & -{g'v_1\over\sqrt{2}} & +{g'v_2\over\sqrt{2}} \cr
  0 & {M_2} & +{gv_1\over\sqrt{2}} & -{gv_2\over\sqrt{2}} \cr
  -{g'v_1\over\sqrt{2}} & +{gv_1\over\sqrt{2}} & 0 & -\mu \cr
  +{g'v_2\over\sqrt{2}} & -{gv_2\over\sqrt{2}} & -\mu & 0 \cr
  } \right)
\end{eqnarray}
can be diagonalized  to give four neutral Majorana states,
\begin{equation}
  \tilde{\chi}^0_i =
  a_{i1} \tilde{B} + a_{i2} \tilde{W}^3 +
  a_{i3} \tilde{H}^0_1 + a_{i4} \tilde{H}^0_2\label{eq:mix}
\end{equation}
($i=1,2,3,4$) the lightest of which,  $\chi_1^0$ or simply $\chi$, is then the candidate for
the particle making up (at least some of) the dark matter in the universe.

The coefficients in Eq.~(\ref{eq:mix}) are normalized such that
for the neutralino
\beq
\sum_{j=1}^4|a_{1j}|^2=1.
\eeq
The properties of the neutralino are quite different depending
on whether is consists mainly of gaugino ($j=1,2$) or higgsino ($j=3,4$)
components. It is therefore customary to define a parameter, $Z_g$,
which tells the size of the gaugino fraction:
\beq
Z_g=\sum_{j=1}^2|a_{1j}|^2.
\eeq
A neutralino is said to be gaugino-like if $Z_g\gsim 0.99$,
higgsino-like if $Z_g\lsim 0.01$, and mixed otherwise.

For simplicity, one often makes a
diagonal ansatz for the
 soft supersymmetry-breaking parameters in the sfermion sector.
This allows the squark mass matrices to be diagonalized analytically.
Such an ansatz implies the absence of
tree-level flavor changing neutral currents (FCNC) in all sectors of the
model. In models inspired by low-energy supergravity with
a universal scalar mass at the grand-unification (or Planck) scale
 the running of the scalar masses down to the electroweak scale
generates off-diagonal terms and tree-level FCNC's in the squark
sector. (For a discussion of this class of models, and of effects
related to relaxing the assumption of universal scalar masses, see
\cite{ellis}.) In most of the estimates of detection rates given
below, we will adhere to a purely phenomenological approach, where the
simplest unification and scalar sector constraints are assumed, and
no CP violating phases outside those of the Standard Model, but no
supergravity relations are used. This reduces the number of free parameters
to be scanned over in numerical calculations to 7: $\tan\beta$,
$M_1$, $\mu$, $m_A$, and three parameters related to the sfermion
sector (the exact values of the latter are usually not very important in
most of our applications).
In fact, on can reduce the number of parameters further by choosing,
e.g., explict supergravity models, but this only corresponds to
a restriction to a subspace of our larger scan of parameter space.

When using the minimal supersymmetric standard model
in calculations
 of relic dark matter density, one should make sure that all
accelerator  constraints on supersymmetric particles and couplings are
imposed. In addition to the significant restrictions on parameters given by
LEP (e.g., \cite{lepbounds,blondel}), the measurement of the \bsg\ process
is providing important bounds \cite{nath,drees,bsg}, since
supersymmetric virtual particles
may contribute  significantly to this loop-induced decay. These
bounds are also included
in the following analysis.

The relic density calculation in the MSSM for a given set of
parameters is nowadays accurate to a few percent or so. A recent
important improvement is the inclusion of
coannihilations, which can change the relic abundance by a large
factor in some instances \cite{coann,ellisco}.

\begin{figure}[!htb]\begin{center}
\epsfig{file=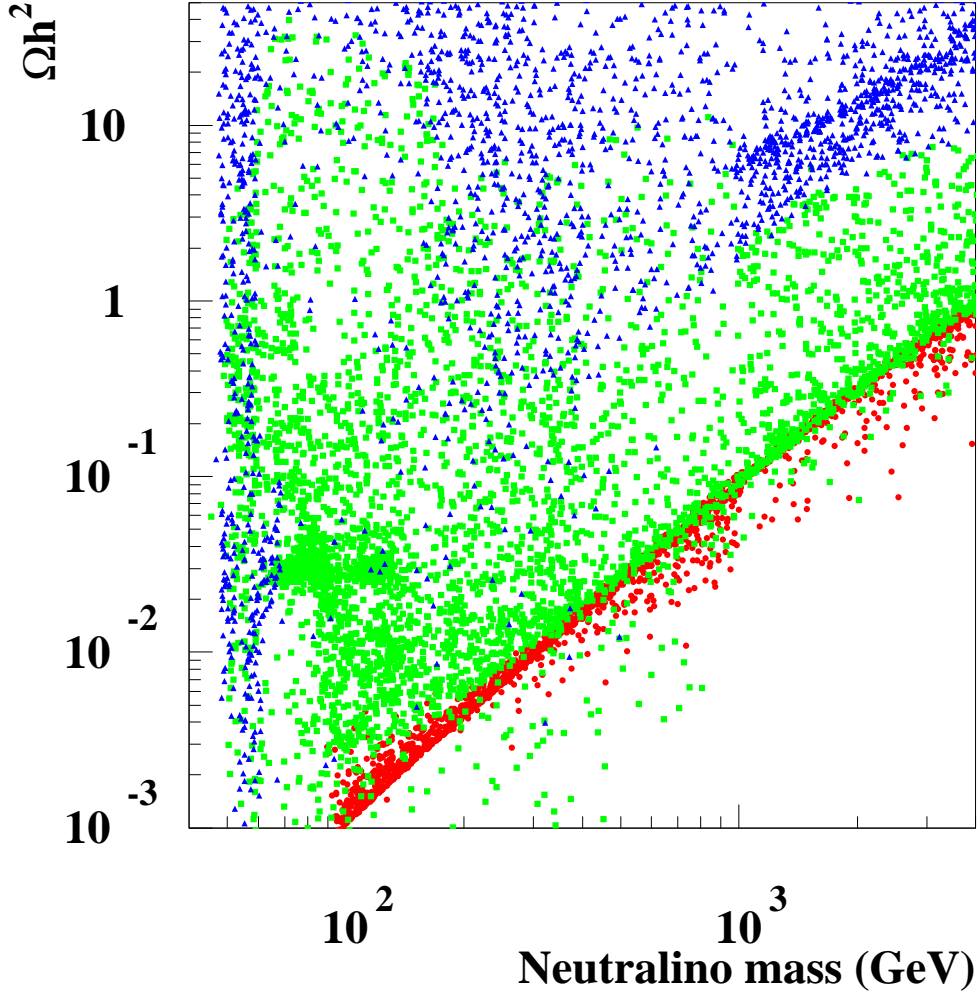,width=\textwidth}
\end{center}
\caption{The computed relic density $\Omega_\chi h^2$ of the lightest
neutralino versus mass in a large sampling of supersymmetric
parameter space. 
The filled circles denote higgsino-like 
models with gaugino fraction $Z_g < 0.01$,
squares are mixed models with  $0.01 < Z_g < 0.99$ and triangles are
gaugino  models
with $Z_g > 0.99$.
For details of the calculation,
see \protect\cite{clumpy,dkpop} and references therein.}
\label{fig:oh2vsmx}
\end{figure}

In Fig.~\ref{fig:oh2vsmx} we show the calculated values of $\Omega_\chi h^2$
versus mass for a large sampling of the 7-dimensional supersymmetric
parameter space (decribed more in detail in \cite{dkpop} and
references therein). As can be seen, there is a very large range
of $\Omega_\chi h^2 $ possible, but the interesting range between, say,
0.1 and 0.2 (assuming $h\sim 0.7$) is comfortably reached by a large
set of models, as anticipated in the discussion after
Eq.~(\ref{eq:relicdensity}). It is important to notice that the
structures that can be seen in the Figure are caused mainly by
the way the supersymmetric parameter space has been sampled. For
instance, the almost straight diagonal band on the lower side
is made up of models which are nearly pure higgsinos. By making a
cut on the maximal higgsino fraction this band would disappear.

\subsection{Other supersymmetric candidates}

Although the neutralino is considered by most workers in the
field to be the preferred supersymmetric dark matter candidate,
we mention briefly here also some other options.
One possibility
is provided by the sneutrino, the supersymmetric partner of one
of the neutrinos. This would give rise to a large coupling to
ordinary matter (and also annihilation cross section) through
$Z$ boson exchange, and has therefore long been thought to be
disfavoured \cite{ellis0}. However, in non-minimal models these
$Z$ couplings may be suppressed, and there are versions which
are still viable, although rather constrained \cite{murayama}.

If the axion exists, and if the underlying theory is supersymmetric,
there should also exist a spin-1/2 partner, the axino. If this
is the lightest supersymmetric particle and is in the multi-GeV
mass range, it could compose the cold dark matter of the universe
\cite{covi}. If it is lighter, it could act as mixed dark matter
\cite{bonometto}
with a non-thermal component arising from
neutralino decay into axinos and
perhaps mix with neutrinos if $R$-parity is violated
\cite{goto}.

A completely different type of supersymmetric dark matter candidate
is provided by so-called Q-balls \cite{kusenko}, non-topological
solitons predicted to be present in many versions of the
theory. These are produced in a non-thermal way and may have large
lepton or baryon number. They could produce unusual ionization signals
in neutrino telescopes, for example. However, the unknown properties
of their precise
formation mechanism means that their relic density may be far below
the level of observability, and a value around the observationally
favoured $\Omega_M\sim 0.3$ may seem fortuitous.

Of course, there remains the possibility of dark matter being
non-supersymmetric WIMPs. However, for the general reasons explained
in Section~\ref{sec:relic}, the interaction cross sections should then
be quite similar as for supersymmetric particles. Since, the rates
in the MSSM are completely calculable once the supersymmetry parameters
are fixed, these particles, in particular neutralinos, serve
as important templates for reasonable dark matter candidates when
it comes to designing experiments with the purpose of  detecting
dark matter WIMPs.

\section{Detection methods for neutralino dark matter}\label{sec:detection}

The ideal situation would appear if supersymmetry were
discovered at accelerators, so that direct measurements of the
mass of the lightest supersymmetric particle, its couplings and
other properties could be performed. This would give a way
to check from very basic principle if this particle is a good
dark matter candidate - if it is electrically neutral and has the
appropriate mass and couplings to give the required relic density
to provide $\Omega_M\sim 0.3$. So far, no signal of supersymmetry
has been found at either LEP or Fermilab, but hopefully
the situation may change as Fermilab's new Main Injector gets
into operation. It may be, however, that one will have
to wait for CERNs Large Hadron Collider to come into operation some time
after 2005 before a signal of supersymmetry may be seen.
(An indirect piece of evidence for supersymmetry would be the discovery
of a Higgs particle below around 130 GeV. In the non-supersymmetric Standard
Model the Higgs could be much heavier.)

The long time scale of
designing and building new accelerators (and the possibility that
the dark matter particles may be extremely heavy or otherwise
difficult to produce and detect) means that it is certainly
worthwhile to explore the possibility to detect the putative dark
matter particles as they move in the Galactic halo. This can be done
directly in terrestrial detectors sensitive to
the nuclear recoil and/or ionization caused by the passing wind of
dark matter particles, or indirectly by detecting products of annihilations
of dark matter particles such as gamma rays, antiprotons or positrons,
in the Galactic halo or in the Earth or Sun (in
which case neutrinos could give an indirect signal).

\subsection{Direct searches of halo dark
matter}\label{subs:direct}

As explained above, we use the neutralino of the MSSM as a template
for an excellent dark matter candidate.
If these neutralinos are indeed the CDM needed on galaxy scales and larger,
there should be a substantial flux of these particles in the Milky
Way halo. Since the interaction strength  is
essentially given by the same weak couplings as, e.g., for neutrinos
there is a non-negligible chance of detecting them in low-background
counting experiments \cite{goodmanwitten}.
Due to the large parameter space of MSSM, even
with the simplifying assumptions above, there is a rather wide span of
predictions for the event rate in detectors of various types. It is
interesting, however, that the models giving the largest rates are
already starting to be ruled out by present direct detection
experiments \cite{bsg,bottino}.

If we assume a local neutralino halo
density of $\rho_\chi=
\rho_\odot\sim 0.3$ GeV/cm$^{3}$, and a typical galactic velocity
of neutralinos of $v/c\sim 10^{-3}$, the flux of particles of mass 100 GeV at
the location of a detector at the Earth is roughly $10^{9}$
m$^{-2}$\,s$^{-1}$. Although this may seem as a high flux, the
interaction rate has to be quite small, since we saw in
Section~\ref{sec:relic} that the correct magnitude
of $\Omega_\chi\sim 0.3$ is only achieved if the annihilation
cross section, and therefore by expected crossing symmetry also
the scattering cross section, is of weak interaction strength.

The rate for direct detection of galactic neutralinos, integrated over
deposited energy assuming no energy threshold, is
\begin{equation}
  R = \sum_i N_i n_\chi \langle \sigma_{i\chi} v \rangle ,
\end{equation}
where $ N_i $ is the number of nuclei of species $i$ in the detector,
$n_\chi$ is the local galactic neutralino number density, $
\sigma_{i\chi} $ is the neutralino-nucleus elastic cross section, and
the angular brackets denote an average over $ v $, the neutralino
speed relative to the detector.

The most important direct detection process is elastic scattering on nuclei,
although inelastic processes \cite{ellisinel,aminne,recent} and
scattering on electrons \cite{spergel_el} have also been suggested
in the literature. Since neutralinos are Majorana particles, there
exist selection rules which dictate the form of the effective
interaction Lagrangian. For instance, since a Majorana fermion can
carry no non-zero conserved additive quantum number (due to the
requirement that they be self-charge-conjugate fields), the vector
current vanishes identically. The most important non-vanishing
currents are then the scalar-scalar coupling giving a spin-independent
effective interaction, and the axial-axial  current which couples
proportionally to the spin of the nucleus:
\beq
{\cal L}_{\rm eff}=f_{SI}\left(\bar\chi\chi\right)\left(\bar NN\right)
+f_{SD}\left(\bar\chi\gamma^\mu
\gamma^5\chi\right)\left(\bar N\gamma_\mu\gamma^5 N\right).
\eeq
Usually, it is the spin-independent interaction that gives the
most
important contribution in realistic target materials (such as Na, Cs,
Ge, I, or Xe), due to the enhancement caused by the coherence of
all nucleons in the target nucleus.

The neutralino-nucleus elastic cross section can be written as
\begin{equation}
  \sigma_{i\chi} = {1 \over 4 \pi v^2 } \int_{0}^{4 m^2_{i\chi} v^2}
  \mbox{\rm d} q^2 G_{i\chi}^2(q^2) ,
\end{equation}
where $ m_{i\chi} $ is the neutralino-nucleus reduced mass, $q$ is the
momentum transfer and $G_{i\chi}(q^2) $ is the effective
neutralino-nucleus vertex. One may write
\begin{equation}
  G^2_{i\chi}(q^2) = A_i^2 F^2_{SI}(q^2) G_{SI}^2 +
  4 \lambda_i^2 J(J+1) F^2_{SD}(q^2) G_{SD}^2 ,
  \label{detrate1}
\end{equation}
which shows the coherent enhancement factor $A_i^2$ for the
spin-independent cross section. A reasonable approximation for the
gaussian scalar and axial nuclear form factors is \cite{Gould87}
\begin{equation}
  F_{SI}(q^2) = F_{SD}(q^2) =
  \exp(-q^2R_i^2/6\hbar^2) ,
\end{equation}
\begin{equation}
  R_i = ( 0.3 + 0.89 A_i^{1/3} )\  {\rm fm} ,
\end{equation}
which gives good approximation to the integrated
detection rate \cite{ellisflores} (but is less accurate for
the differential rate \cite{engel}). Here $\lambda_i$ is related to
the average spin of the nucleons making up the nucleus. For the relation
between $G_{SI}$, $G_{SD}$ and $f_{SI}$, $f_{SD}$ as well as a discussion of
the several Feynman diagrams which contribute to these
couplings, see e.g.~\cite{bsg,nojiri,bailin}.
One should be aware that both the choice of nuclear form factors and
effective neutralino-nucleon vertices as well as the numerical values adopted for
the nucleon matrix elements are at best approximate. A more
sophisticated treatment (see discussion and references in \cite{jkg})
would, however, change the values by much less than the spread due to
the unknown supersymmetric parameters.

For a target consisting of $N_i$ nuclei the differential
scattering rate per unit time and unit recoil energy $E_R$
is given by
\beq
S_0(E_R) =
{dR\over dE_R}=N_i{\rho_\chi\over m_\chi}
\int \,d^3v\,f(\vec v)\,v {d\sigma_{i\chi}\over dE_R}(v,E_R). \label{eq:diffscatt}
\eeq
The nuclear recoil energy $E_R$ is given by
\beq
E_R={m_{i\chi}^2v^2(1-\cos \theta^*)\over {m_i}}
\eeq
where $\theta^*$ is the scattering
angle in the center of mass frame.
The range and slope of the recoil energy spectrum is essentially
given by non-relativistic kinematics. For a low-mass $\chi$, the spectrum is
steeply falling with $E_R$; interaction with a high-mass $\chi$
gives a flatter spectrum with higher cutoff in $E_R$.

\begin{figure}[!htb]\begin{center}
\epsfig{file=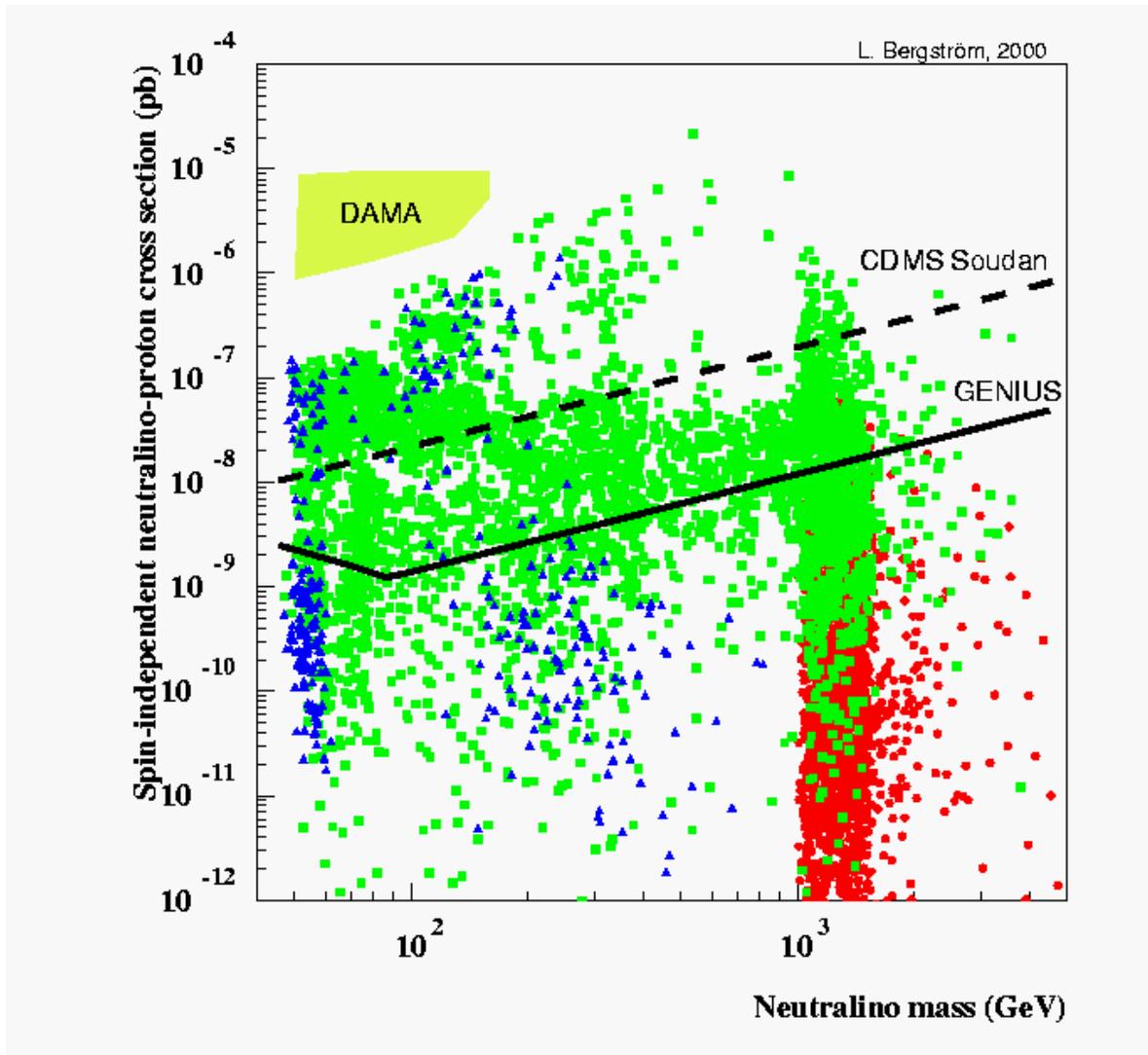,width=\textwidth}
\end{center}
\caption{The neutralino-proton cross section versus mass for
a large scan in supersymmetric parameter space fulfilling
$0.1 < \Omega_\chi h^2 < 0.2$, $m_{\chi^\pm}>95$ GeV and
$m_{H_2^0}> 100$ GeV (the cross section on neutrons is very
similar). The filled circles denote higgsino-like 
models with gaugino fraction $Z_g < 0.01$,
squares are mixed models with  $0.01 < Z_g < 0.99$ and triangles are
gaugino  models
with $Z_g > 0.99$. The region consistent with the claimed DAMA signal
\protect\cite{dama} is indicated. The approximate 
expected  sensitivity curves
of the proposed CDMS-Soudan and GENIUS experiments are also shown.}
\label{fig:sigmapvsmx}
\end{figure}

The total predicted rate integrated over recoil energy above a given
generally (detector-dependent)
threshold can be compared with upper limits coming from various
direct detection experiments. In this way, limits on the
$\chi$-nucleon cross section have been obtained as a function of the mass
 $m_\chi$
 \cite{directlimits}. The cross section on neutrons is usually
very similar to that on protons, so in general only the latter
is displayed. In Fig.~\ref{fig:sigmapvsmx} is shown a
scatter plot of the spin-independent neutralino-proton cross
 sections as a function of neutralino mass predicted in an extensive scan
of the MSSM parameter space. This is the  same scan as that used
in \cite{dkpop} (see also references therein) except for updated
LEP bounds: $m_{\chi^\pm}>95$ GeV, $m_{H_2}>100$ GeV. These latter bounds
are simplified and somewhat overconstraining. The actual bounds depend
on other parameters such
as $\tan\beta$ and the detailed decay modes.
In this figure, only the range $0.1 < \Omega_\chi h^2 <0.2$
has been selected, as this is the favoured range for CDM as
discussed before. This more lower bound is more restrictive than usually
employed ($\Omega h^2 > 0.025$), but also more strongly motivated by
cosmology. It has the effect of excluding some model with large
scattering cross sections. Also shown are the present experimental upper bounds as well
as the region consistent with the possible DAMA signal discussed
below. 
A major step forward is expected within a couple of years.
For example, the CDMS experiment will be moved from a shallow
Stanford site to the well-shielded Soudan mine. This together
with a  larger detector mass and other improvements will enable
a thorough search well beyond the range suggested by DAMA.
Also in Europe there are several ambitious endeavours underway,
such as the GENIUS detector \cite{genius}, CRESST2 \cite{cresst}
and UKDMC \cite{ukdmc}.
As can be seen in the Figure, some of these experiments will start to
probe interesting regions of neutralino dark matter.

The  rate in Eq.~(\ref{eq:diffscatt}) is strongly dependent
on the velocity $v$ of the neutralino with respect to the target
nucleus. Therefore, as explained in Section~\ref{subs:halos}
an annual modulation of the counting rate
is in principle possible, due to the motion of the Earth around
the Sun \cite{drukier}. One can thus write
\beq
S(E_R,t) = S_0 (E_R) + S_m (E_R) \cos \left[\omega (t-t_0)\right],
\eeq
where $\omega = 2\pi/365$ days$^{-1}$. Starting to count time
in days from January 1$^{\rm st}$, the phase  is $t_0 = 153$ days
since the maximal signal occurs when the direction of
motion of the Earth
around the Sun and the Sun around the galactic center coincide maximally,
which happens on June 2$^{\rm nd}$ every year \cite{drukier}. Similarly, the
counting rate is expected to be the lowest December 2$^{\rm nd}$ every
year. Here $S_0 (E_r)$ is the average
 differential scattering rate  in
Eq.~(\ref{eq:diffscatt})and $S_m (E_R)$ is the
modulation amplitude of the rate.
The relative size of $S_m (E_R)$ and $S_0 (E_R)$
depends on the target and neutralino mass as well as on $E_R$. Typically $S_m (E_R)$
is of the order of a few percent of $S_0 (E_R)$, but
may approach 10 \%  for small $m_\chi$ (below, say, 50 GeV) and small $E_R$
(below some 10 keV).

Since the basic couplings in the MSSM are between neutralinos and
quarks, there are uncertainties related to the hadronic physics
step which relates quarks/gluons with nucleons as well the step from
nucleons to nuclei. These uncertainties are substantial, and can
plague all estimates of scattering rates by at least a factor of
2,
maybe even by an order of magnitude \cite{bottinonew}.
The largest rates, which as first shown in \cite{bsg} could be
already ruled out by current experiments, are generally obtained
for mixed neutralinos, i.e. with $Z_g$ neither very near 0 nor very near 1,
and for relatively light Higgs masses (since Higgs bosons mediate
a scalar, spin-independent exchange interaction). This means that
the LEP bounds on the supersymmetric Higgs particle masses put
relevant constraints on the predicted detection rates.

The experimental situation is becoming interesting as several
direct detection experiments after many years of continuing
sophistication are starting to probe interesting parts of
the parameter space of the MSSM, given reasonable, central values
of the astrophysical and nuclear physics parameters. Perhaps
most striking is the evidence for an annual modulation
effect claimed to be seen in the NaI experiment DAMA \cite{dama}. This
has been interpreted as possibly being due to a neutralino of
the MSSM \cite{damabott,nath2}. It seems premature, however, to
draw strong conclusions from this experiment alone. Besides
some cloudy  experimental issues \cite{tao,cdms}, the implied
scattering rate seems somewhat too high for the MSSM, given
the recent strong Higgs mass bounds from LEP operating above 200
GeV (especially if one wants $\Omega_\chi\gsim 0.2$),
unless one really stretches
the astrophysical \cite{bottinoast,brl} and nuclear
physics quantities \cite{bottinonew}. This is seen in
Fig.~\ref{fig:sigmapvsmx}, where the DAMA region is not
populated by any points. The main reason for this is
the $\Omega$ cut. If the lower bound is relaxed to $\Omega_\chi h^2>0.025$,
some models would appear in the DAMA region. It is probably also possible,
but not very attractive,
to find points with higher rates by making special samplings 
in parameter space designed just to find such high rates.
Clearly, more sensitive
experiments are needed to settle this issue.

Many of the present day detectors are severely hampered by a large
background of various types of ambient radioactivity or cosmic-ray induced
activity (neutrons are a particularly severe problem since
they may produce recoils which are very similar to the expected
signal). A great improvement in sensitivity would be acquired if
one could use directional information about the recoils \cite{spergel_dir}.
There
are some very interesting developments also along this line
\cite{dan}, but a full-scale detector is yet to be built.
Direction-sensitive detectors would have an even bigger
advantage over pure counting experiments if the dark matter
velocity distribution
is less trivial than the commonly assumed maxwellian, as has been
recently suggested \cite{sikivie,damourkrauss}.

\subsection{Indirect searches}

Besides these possibilities of direct detection of supersymmetric dark
 matter
 (with even a
 weak indication of the existence of a  signal \cite{dama}),
 one also has the possibility of indirect detection through
 neutralino annihilation
in the galactic halo. This is becoming a promising method thanks
 to very powerful new detectors for
cosmic gamma rays  and neutrinos planned and under construction.

There has  been a  balloon-borne  detection experiment
\cite{Barwick},
with increased sensitivity to eventual positrons from neutralino annihilation,
where an excess of positrons over that expected from ordinary sources
was found. However, since there are many other possibilities to
create positrons by astrophysical sources, e.g., near the centre of
the Milky Way, the interpretation is not yet conclusive. Also, another
measurement does not confirm this excess \cite{mirko}.

 Antiprotons, $\bar p$,
from neutralino annihilations were long hoped to give a useful
signal \cite{antiprotons}, and there have been several balloon-borne
experiments \cite{caprice,bess} performed
and a very ambitious space experiment, AMS,
to search for antimatter is under way \cite{ting}.
For kinematical reasons, antiprotons
created by pair-production in cosmic ray collisions with interstellar
gas and dust are born with relatively high energy, whereas
antiprotons from neutralino annihilation populate also the sub-100
MeV energy band.

However, it was found recently \cite{pbar,tomg} that the cosmic-ray
induced antiprotons may populate also the low-energy region
to a greater extent than previously thought, making the extraction
of an eventual supersymmetric signal much more difficult. There are
basically three effects which cause this problem. First, helium
and other heavier elements in the interstellar medium give an antiproton
yield at lower energy than hydrogen, since the centre of mass system
is kinematically closer to the galactic rest frame. Secondly, antiprotons
may be produced at incident and outgoing nominal kinetic energies
below the threshold energies valid for proton-proton collisions,
due to collective nuclear effects. And, thirdly, secondary elastic and
inelastic interactions
of produced antiprotons give a ``tertiary'' $\bar p$ component at lower
kinetic energy.

Another
 problem that plagues estimates of the signal strength of both positrons and
antiprotons is  the uncertainty of the galactic propagation model
and solar wind
 modulation.

 Even allowing for large such systematic effects, the
 measured antiproton flux gives, however, rather stringent limits on
MSSM models with the highest annihilation rates. One can also
use the experimental upper limits to bound from below the
 lifetime of hypothetical $R$-parity violating decaying neutralinos
 \cite{baltz}. There may in some scenarios with a clumpy halo
(which enhances the annihilation rate) be a possibility to detect
heavy neutralinos through spectral features above several GeV \cite{piero}.

A very rare process in proton-proton collisions, antideuteron production,
may be less rare in neutralino annihilation \cite{fiorenza}. However,
the fluxes are so small that the possibility of detection seems marginal
even in the AMS experiment.
\subsection{Indirect detection by gamma rays from the halo}

With the problem of a lack of clear signature of positrons and antiprotons, one would expect that
the situation
 of gamma rays and neutrinos is similar, if they only arise from
secondary decays in the
annihilation process. For instance, the gamma ray spectrum arising from
the fragmentation of fermion and gauge boson final states is quite
featureless and gives the bulk of the gamma rays at low energy where the
cosmic gamma ray background is severe. However, an advantage is the directional
information that photons carry in contrast to charged particles which
random walk through the magnetic fields of the Galaxy \cite{contgammas}.

\subsubsection{Gamma ray lines}\label{subs:lines}

An early idea was to look for a spectral feature, a line, in the
radiative annihilation process to a charm-anticharm bound state
$\chi\chi\to (\bar c c)_{\rm bound} +\gamma$ \cite{STS}.
However, as the experimental lower bound on the lightest neutralino
became higher it was shown that form factor suppression rapidly
makes this process unfeasible \cite{berg-snell}. The  surprising
discovery was made that
the loop-induced annihilations
$\chi\chi\to\gamma\gamma$ \cite{berg-snell,gammaline} and $\chi\chi\to Z\gamma$
\cite{zgamma} do not suffer at all from any form factor suppression (this was
subsequently shown to be related to the famous triangle anomaly of quantum
field theory \cite{rudaz}).

The rates of these processes are difficult to estimate because of
uncertainties in
the supersymmetric parameters, cross sections and halo density profile. However,
in contrast to the other proposed detection methods they have
the virtue of giving  very
distinct, ``smoking gun'' signals of
monoenergetic photons with energy $E_\gamma = m_\chi$
(for $\chi\chi\to\gamma\gamma$) or $E_\gamma = m_\chi
(1-m_{Z}^2/4m_{\chi}^2)$ (for $\chi\chi\to Z\gamma$)
emanating from annihilations in the halo.

The detection probability of a gamma ray signal, either continuous or line, will of course depend
sensitively on the density profile of the dark matter halo.
To illustrate this point, let us consider the characteristic
angular dependence of
the gamma-ray line intensity from neutralino annihilation $\chi\chi\to\gamma\gamma$
in the galactic halo.
Annihilation of neutralinos in an isothermal halo
with core radius
$a$ leads to a gamma-ray flux along the line-of-sight direction $\hat n$ of
\begin{eqnarray}
     {d{\cal F} \over {d \Omega}}\left(\hat n \right)\simeq \,
     (2\times10^{-13} {\rm cm}^{-2} {\rm s}^{-1} {\rm sr}^{-1})\times &
     &\nonumber\\
    \left({\sigma_{\gamma\gamma} v\over 10^{-29}\ {\rm cm}^{-3}{\rm
    s}^{-1}}\right)
     \left({\rho_\chi\over 0.3\ {\rm GeV}\,{\rm cm}^{-3}}\right)^2
     \left({100\,{\rm GeV}\over m_\chi}\right)^2 \left({R\over 8.5\ {\rm kpc}}\right)J(\hat n)&&
\end{eqnarray}
where
 $\sigma_{\gamma\gamma}
v$ is the annihilation rate,
$\rho_\chi$ is the
local neutralino halo density
and $R$ is the distance
to the galactic center.
The integral $J(\hat n)$ is given by
\beq
J(\hat n)={1\over R\rho_\chi^2}\int_{\rm
line-of-sight}\rho^2(\ell)d\ell(\hat n),\label{eq:j}
\eeq
and is evidently very  sensitive to local density variations along the
line-of-sight path of integration. In the case of a
smooth halo, its value ranges from a few at
high galactic latitudes to several thousand for a small angle
average towards the galactic center in the NFW model \cite{BUB}.

 We remind of the
fact that since the
neutralino velocities in the halo are of the order of 10$^{-3}$ of the
velocity of light, the
annihilation can be considered to be at rest. The resulting gamma ray
spectrum is a line
at $E_\gamma=m_\chi$ of relative linewidth 10$^{-3}$ which in
favourable cases
will stand out against background.

The calculation of the $\chi\chi\to\gamma\gamma$ cross section is
technically quite
involved with a large number of loop diagrams contributing. In fact,
only very recently a
full calculation in the MSSM
was performed \cite{newgamma}.
Since the
different contributions all have to be added coherently, there may be
cancellations or
enhancements, depending on the supersymmetric parameters.
The process $\chi\chi\to Z\gamma$
is treated analogously  and has a  similar rate
\cite{zgamma}.

An important contribution, especially for neutralinos that contain
a fair fraction of a higgsino component, is from virtual $W^+W^-$
intermediate states.
This is  true both for the $\gamma\gamma$ and $Z\gamma$ final
state for very massive neutralinos \cite{zgamma}. In fact, thanks to
the effects of coannihilations \cite{coann}, neutralinos as heavy as
several TeV are allowed without giving a too large $\Omega$. These
extremely heavy dark matter candidates (which, however,
would require quite a
degree of finetuning in most supersymmetric models)
are predominantly higgsinos
and have a remarkably large branching ratio into the loop-induced
$\gamma\gamma$ and $Z\gamma$ final states (the sum of these can be as
large as 30\%). If there would exist such heavy, stable neutralinos,
the gamma ray line annihilation process may be the only one which could reveal
their existence in the foreseeable  future (since not even LHC would be sensitive
to supersymmetry if the lightest supersymmetric particle weighs several
TeV)\footnote{
Recently, there has been some interest in TeV
neutralinos due to a claim of a possible structure in existing data
\cite{strausz}. It seems, however, that this claim was based on an
erroneous estimate of the acceptance of the experiments \cite{buckley}. Also, the
purported rate is at least 3 or 4  orders of magnitude larger than
what can be obtained in supersymmetric models \cite{newgamma}.}.

To compute $J(\hat n)$ in Eq.~(\ref{eq:j}), a
model of the dark matter halo has to be chosen,
as discussed in Section~\ref{sec:distribution}.
 The universal halo profile found in
 simulations by Navarro, Frenk and White \cite{NFW}
 has a rather significant enhancement $\propto 1/r$
near the halo centre. (In fact, as was seen in Section~\ref{sec:distribution},
other simulations give even steeper central halo profiles.)
If applicable to the Milky Way, this
 would lead to a much enhanced annihilation
rate towards the galactic centre, and also to a very characteristic
angular dependence of the line signal. This would be very beneficial
when discriminating against the  extragalactic $\gamma$
ray background, and Air Cherenkov Telescopes (ACTs) would be eminently
suited to look for these signals since they have an angular
acceptance which is well matched to the angular size of the Galactic central
region where a cusp is likely to be. However, to search for lines the energy resolution
 has to be at the
$10-20$ \% level, which is difficult but possible with present-day technology.

\begin{figure}[!htb]\begin{center}
\epsfig{file=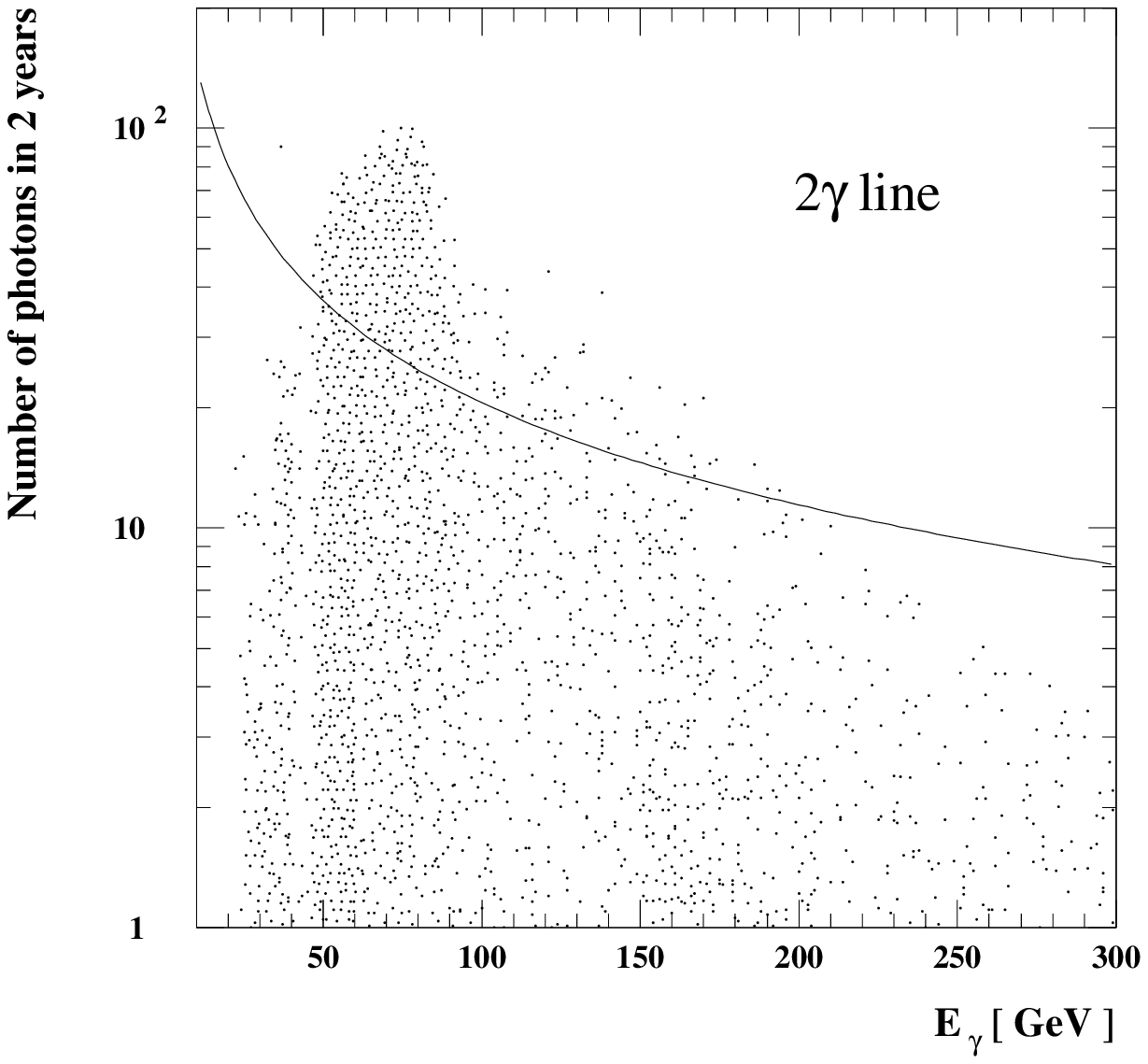,width=7cm}\epsfig{file=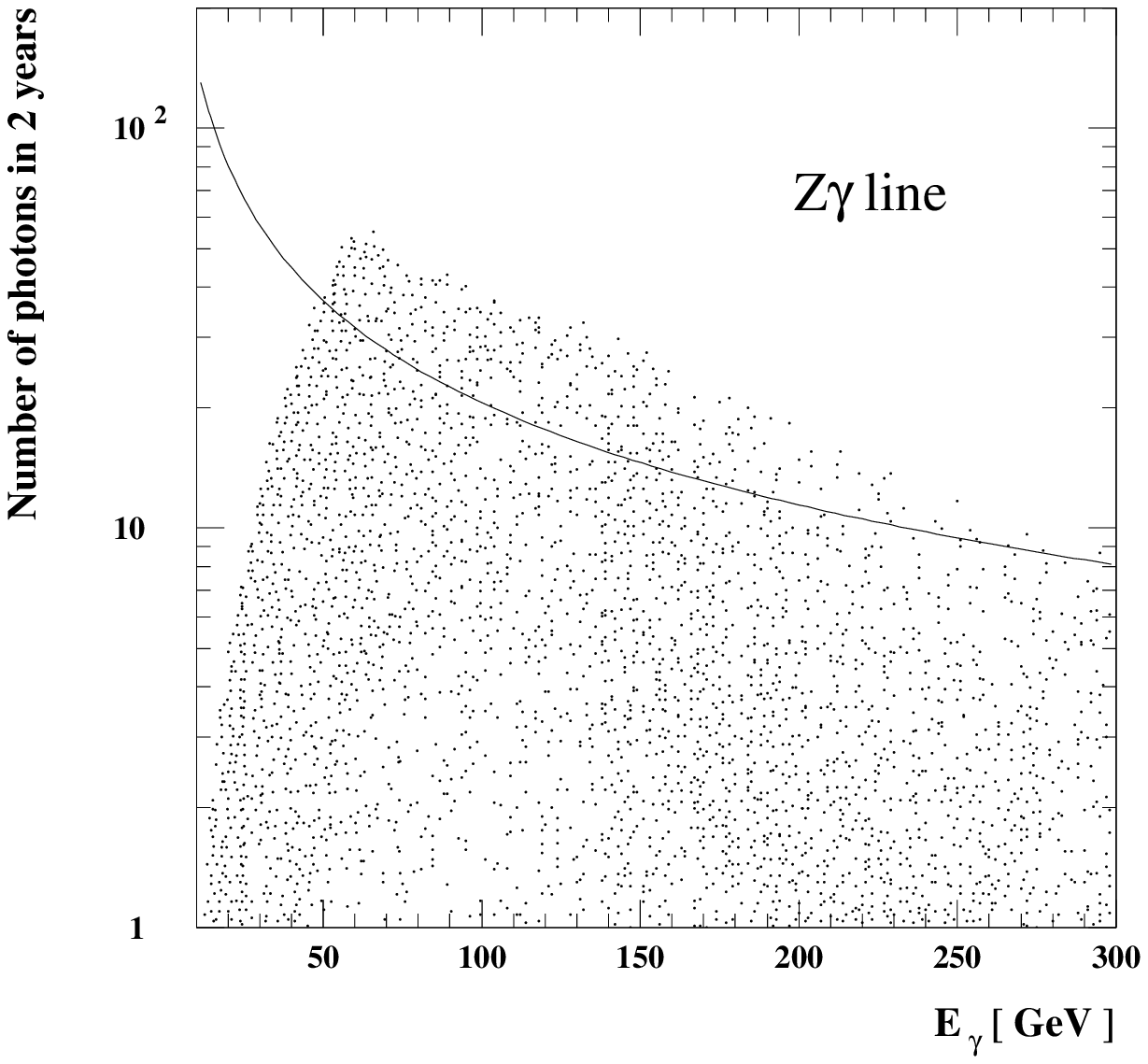,width=7cm}
\end{center}
\caption{Results for the gamma ray line flux from
$\chi\chi\to\gamma\gamma$ (left) and $\chi\chi\to Z\gamma$ (right)
in an extensive
scan of supersymmetric
parameter space in the MSSM \protect\cite{BUB}.
Shown is the number of events versus photon energy in the GLAST
space-borne gamma ray detector
during a 2-year exposure. The
halo profile of \protect\cite{NFW} for the dark matter has been assumed.
The estimated sensitivity of GLAST is shown as the solid line, taking
the diffuse gamma-ray background into account.}
\label{fig:glast}
\end{figure}

Space-borne gamma ray detectors, like the projected GLAST
satellite \cite{glast},
 have a much smaller area (on the order of 1
m$^2$ instead of $10^4-10^5$ m$^2$ for ACTs), but a
correspondingly larger angular acceptance so that the integrated
sensitivity is in fact similar. This is at least true if the Galactic
center does not have a very large dark matter density enhancement which
would favour ACTs. The total rate expected in GLAST can be
computed with much less uncertainty because of the angular
integration. Directional information is obtained and can
be used to discriminate against the diffuse extragalactic
background. A line signal can be searched for with high precision,
since the energy resolution of GLAST will be at the few percent
level. In Fig.~\ref{fig:glast} is shown the potential for GLAST
assuming a NFW profile and an exposure time of 4 years. As
can be seen, there is a region below around 250 GeV where these
processes are observable for the models with the highest rates.

\begin{figure}[!htb]\begin{center}
\epsfig{file=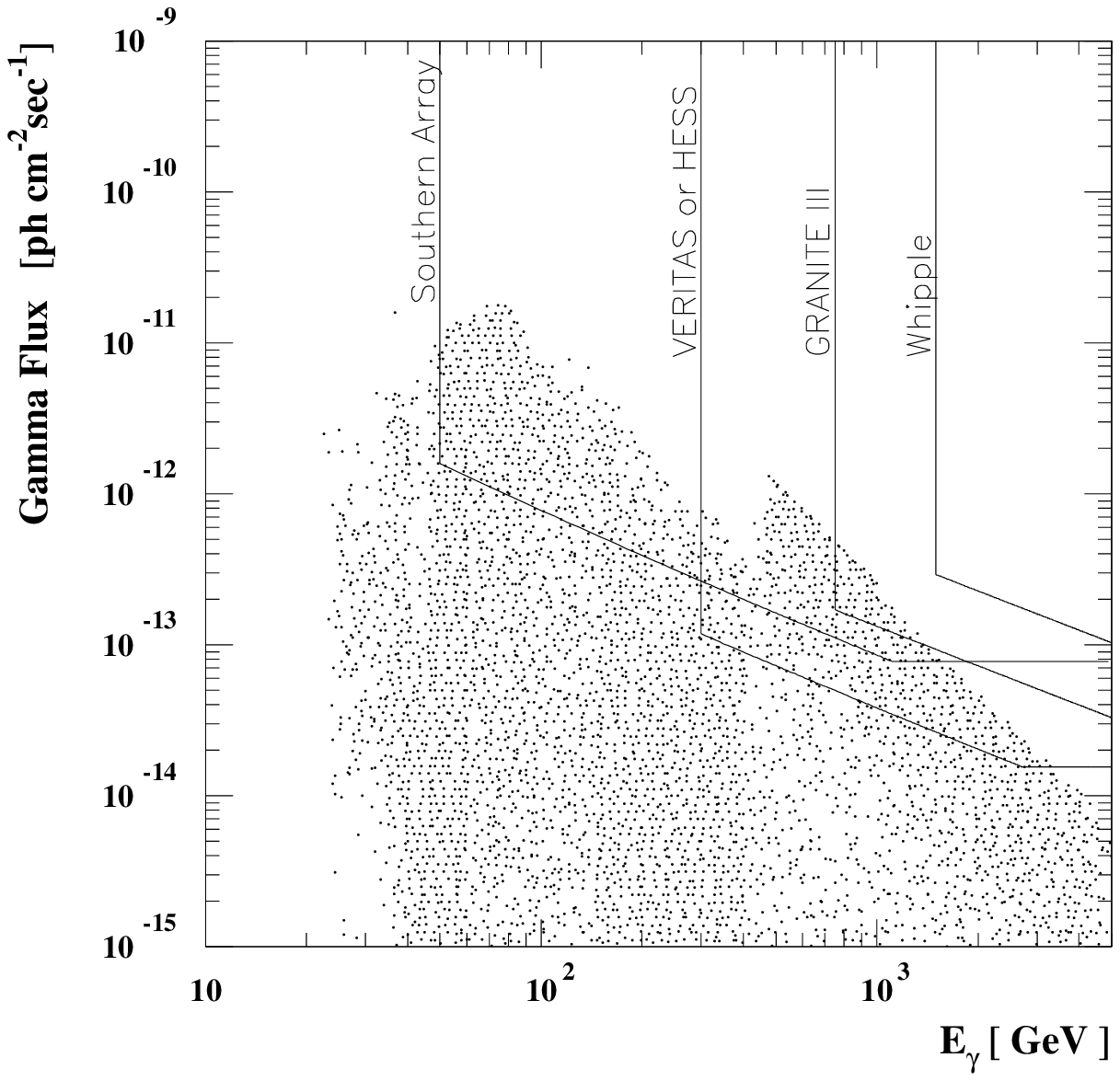,width=7cm}\epsfig{file=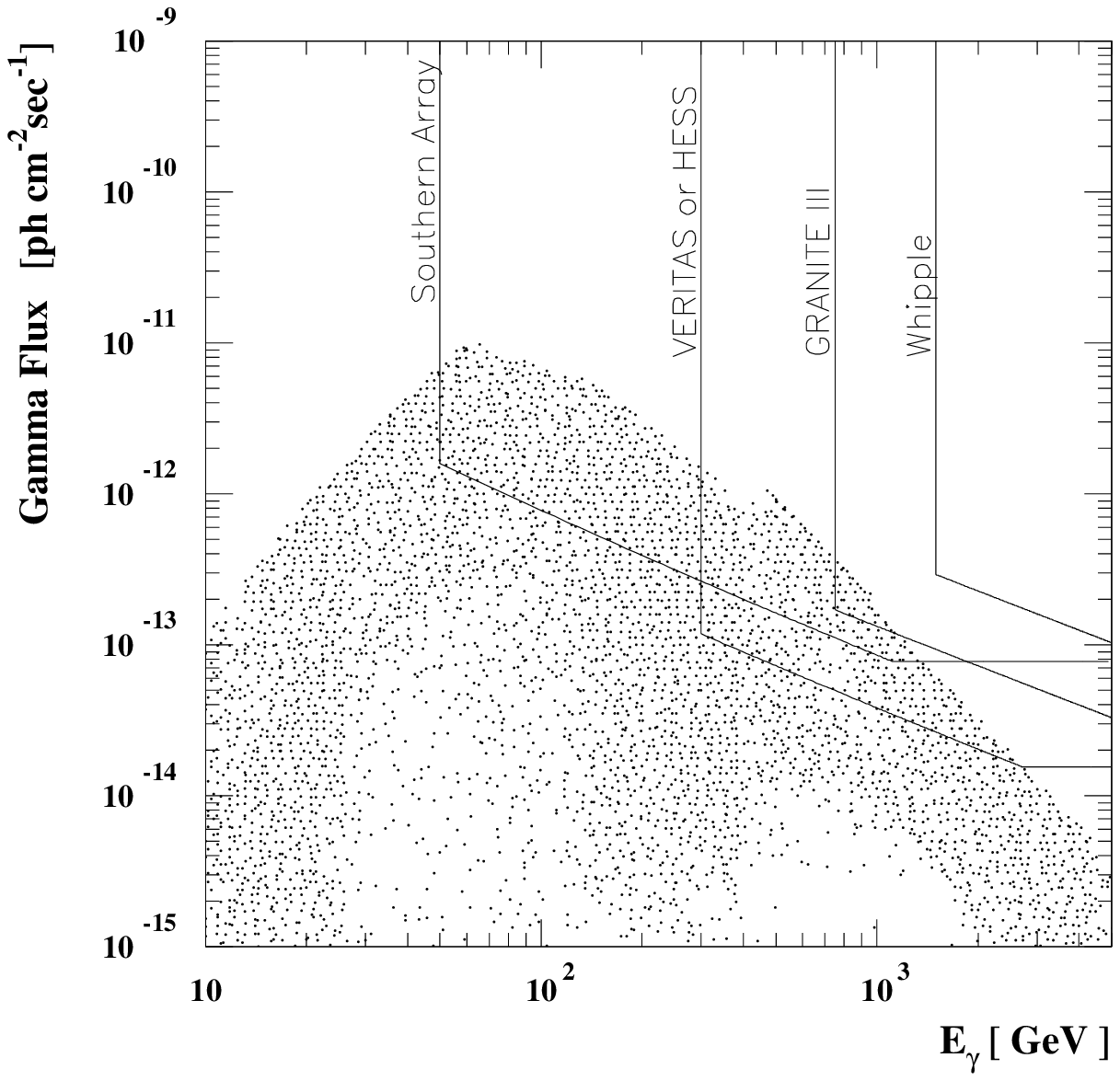,width=7cm}
\end{center}
\caption{Results for the gamma ray line flux from
$\chi\chi\to\gamma\gamma$ (left) and $\chi\chi\to Z\gamma$ (right)
in an extensive
scan of supersymmetric
parameter space in the MSSM \protect\cite{BUB}.
Shown is the number of events versus photon energy in an Air Cherenkov Telescope of area $5\cdot 10^4$
m$^2$  viewing the galactic centre for one year. The
halo profile of \protect\cite{NFW} for the dark matter has been assumed.
}
\label{fig:NC}
\end{figure}

In Fig.\,~\ref{fig:NC} is shown the corresponding gamma ray line flux
in an ACT assuming an effective value of $10^3$ for the average of $J(\hat n)$
over the $10^{-3}$ steradians that typically an
ACT would cover.  (See \cite{BUB} for details.)

\begin{figure}[!htb]\begin{center}
\epsfig{file=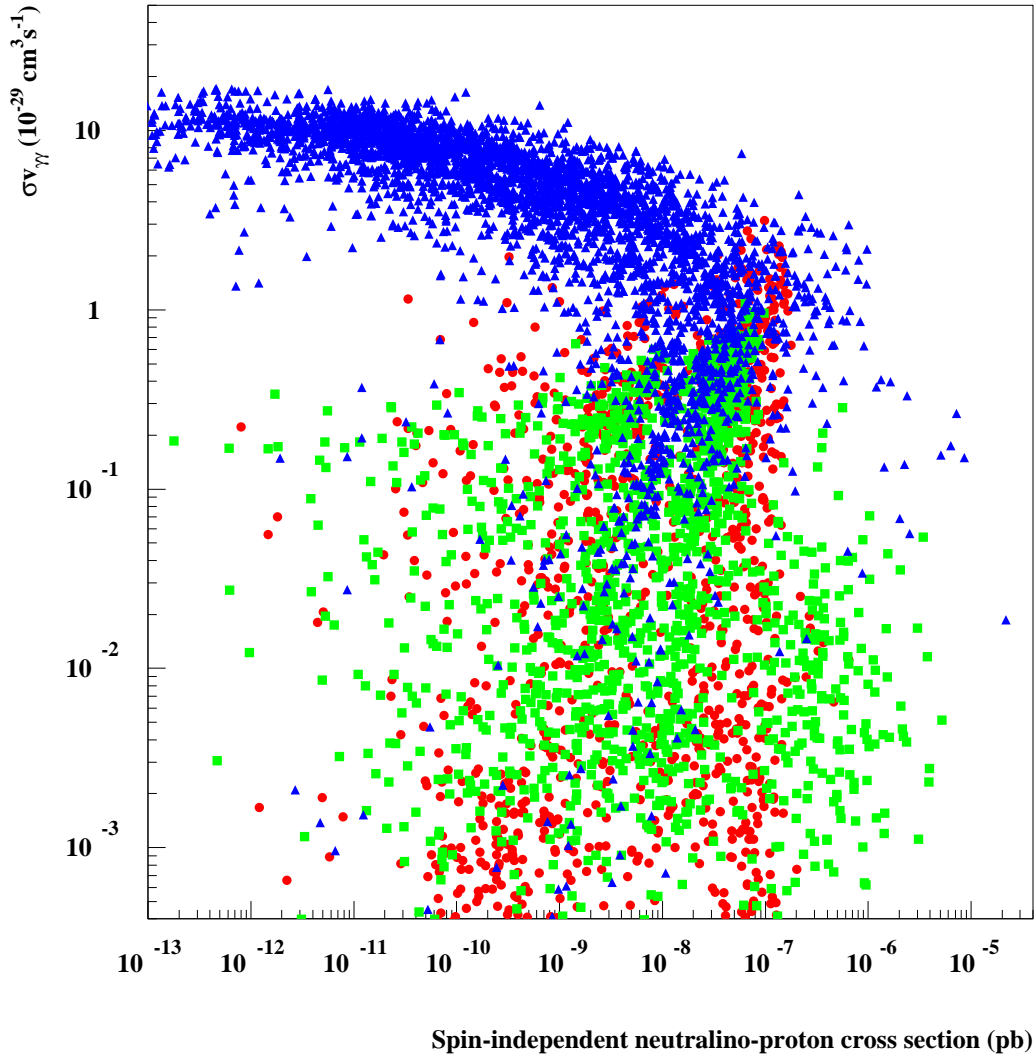,width=\textwidth}
\end{center}
\caption{Comparison of direct/indirect detection strength for the full
sample of points in MSSM parameter space. On the vertical axis
is plotted the value of $\sigma_{\gamma\gamma} v$ (in units
of $10^{-29}$ cm$^3$\,s$^{-1}$),
and on the horizontal axis is shown the neutralino-proton cross section
in picobarns. The filled circles denote models with $m_\chi < 100$ GeV,
squares have $100$ GeV $< m_\chi < 400$ GeV and triangles are models
with $m_\chi > 400$ GeV.
The large spread of values illustrates the complementarity of
the processes. In particular, it is seen how the $\gamma\gamma$ process
has a large rate for the very heavy (higgsino-like) neutralinos which
have a small scattering cross section.
The requirements $m_{\chi^\pm}>95$ GeV, $m_{H_2}>100$ GeV,
 $0.1 < \Omega_\chi h^2 <0.2$ have been imposed.}
\label{fig:comparison}
\end{figure}

It can be seen that the models which give
the highest rates should also be within reach of the new generation of ACTs
presently being constructed. These will have an effective area of almost $10^5$
m$^2$, a threshold of some tens of GeV and an energy resolution
approaching 10 \%. As seen in Fig.\,~\ref{fig:NC}, even TeV masses
 could be accessible through this method. At the low
$m_{\chi}$ end, also a smaller area detector with  better energy
resolution and wider angular acceptance such as the proposed GLAST
satellite could reach discovery potential. It is important to
note that direct detection as discussed in
Section~\ref{subs:direct} and indirect detection through gamma ray
lines are complementary to each other. This is illustrated in
Fig.\,~\ref{fig:comparison}, where
$\sigma_{\gamma\gamma}v$ is displayed against $\sigma_{p\chi}$
for our sampling of the MSSM. As
can be seen, the two types of process are nicely complementing each
other. In particular, models with $m_\chi$ larger than 400 GeV
which generally have a small direct detection rate
usually have a substantial annihilation rate in the $\gamma\gamma$
mode.

\subsubsection{Indirect detection through neutrinos}

The density of
neutralinos in the halo is not large enough to give a measurable flux
of secondary neutrinos, unless the dark matter halo is very clumpy \cite{clumpy}.
In particular, the central Galactic black hole may have interacted with the dissipationless
dark matter of the halo so that a spike of very high dark matter density
may exist right at the Galactic centre \cite{gondolosilk}. However, the existence
of these different forms of density enhancements are very uncertain and depend
extremely sensitively on presently completely unknown aspects of the formation
history of the Milky Way.

More model-independent predictions (where essentially only the relatively
well-determined local halo dark matter density is of importance)
can be made for neutrinos  from the centre
of the Sun or Earth, where
neutralinos may have been gravitationally trapped and therefore their density
enhanced.  As they annihilate, many of the possible final states
(in particular, $\tau^+\tau^-$ lepton pairs, heavy quark-antiquark pairs
and, if kinematically allowed, $W^\pm H^\mp$,
$Z^0H_i^0$, $W^+W^-$ or $Z^0Z^0$ pairs) give
after decays and perhaps hadronization energetic neutrinos which
will propagate out from the interior of the Sun or Earth.
(For neutrinos from the Sun,
energy loss of the hadrons
in the solar medium and the energy loss of neutrinos
have to be considered \cite{ritzseckel,jethesis}). In particular,
the muon neutrinos are useful for indirect detection of
neutralino annihilation processes, since muons
have a quite long range in a suitable detector medium like ice or water.
Therefore they can be detected through their Cherenkov radiation after
having been produced
at or near the detector, through the action of
a charged current weak interaction
$\nu_\mu + A \to \mu + X$.

Detection of neutralino annihilation into neutrinos is
one of the most promising indirect detection methods,
and  will be subject to  extensive experimental investigations in view
of the new neutrino telescopes (AMANDA, IceCube, Baikal, NESTOR, ANTARES)
planned or under construction \cite{halzen}. The advantage shared with
gamma rays is that neutrinos keep their original direction. A high-energy
neutrino signal in the direction of the centre of the Sun or Earth
is therefore an excellent experimental signature which may stand up against
the background of neutrinos generated by cosmic-ray interactions in the
Earth's atmosphere.

The differential neutrino flux from neutralino annihilation is
\beq
\frac{dN_\nu}{dE_\nu} =
\frac{\Gamma_A}{4\pi D^2} \sum_{f}
B^{f}_{\chi}\frac{dN^f_\nu}{dE_\nu}
\eeq
where $\Gamma_A$ is the annihilation rate,
$D$ is the distance of the detector from the source (the
central region of the Earth or the Sun), $f$ is the neutralino pair
annihilation final states,
and $B^{f}_{\chi}$ are the branching ratios into the final state $f$.
 $dN^f_\nu/dE_{\nu}$ are the energy
distributions of  neutrinos generated by the final state $f$.
Detailed calculations of these spectra
can be made using Monte Carlo
methods \cite{bottnuflux,jethesis,BEG}.

The neutrino-induced muon flux may be detected in a neutrino telescope
by measuring the muons that come from the direction of the centre
of the Sun or Earth. For a shallow detector, this usually has to
be done in the case of the Sun by looking (as always the case for
the Earth) at upward-going muons, since there is a huge background
of downward-going muons created by cosmic-ray interactions in the
atmosphere. There is always in addition a more isotropic
background coming from muon neutrinos created on the other side of
the Earth in such cosmic-ray events (and also from cosmic-ray
interactions in the outer regions of the Sun).
The flux of muons at the detector is  given by

\beq
\frac{d N_\mu}{d E_\mu}
= N_A \int^\infty_{E_\mu^{\rm th}} d E_\nu
\int_0^\infty d\lambda \int_{E_\mu}^{E_\nu}
d {E'_\mu }\,\,
P(E_\mu,E'_\mu; \lambda)\,\,
\frac{d \sigma_\nu (E_\nu,E'_\mu)}{d E'_\mu} \,\,
\frac{d N_\nu}{d E_\nu}\, ,
\label{eq:muflux}
\eeq
where $\lambda$ is the muon range in the medium (ice or water
for the large detectors in the ocean or at the South Pole,
or rock which surrounds the smaller underground detectors),
$d \sigma_\nu (E_\nu,E'_\mu) / d E'_\mu$ is
the weak interaction cross section for production of a muon of
energy $E'_\mu$ from a parent neutrino of energy $E_\nu$, and
$P(E_\mu,E'_\mu; \lambda)$ is the
probability for a muon of initial energy $E'_\mu$
to have a final energy $E_\mu$ after passing
 a path--length $\lambda$ inside the detector medium.
$E_\mu^{\rm th}$ is the detector threshold energy, which for
``small''
neutrino telescopes like Baksan, MACRO and Super-Kamiokande is
around 1 GeV.
Large area neutrino telescopes in the ocean  or in Antarctic ice
typically
have thresholds of the order of tens of GeV, which makes them
sensitive mainly to heavy neutralinos (above 100 GeV)
\cite{begnu2}. Convenient approximation formulas relating the observable
muon flux to the neutrino flux at a given energy can be found
in \cite{halzenreview}.

The integrand in Eq.~(\ref{eq:muflux}) is weighted towards high
neutrino energies, both because the cross section $\sigma_\nu$
rises approximately linearly with energy and because the average
muon energy, and therefore the range $\lambda$, also grow
approximately linearly with $E_\nu$. Therefore, final states
which give a hard neutrino spectrum (such as heavy quarks, $\tau$
leptons and $W$ or $Z$ bosons) are usually more important
than the soft spectrum arising from light quarks and gluons.

The rate of change of the number of  neutralinos $N_\chi$ in the Sun or
Earth is governed by the equation
\beq
\dot N_\chi=C_C-C_AN_\chi^2\label{eq:ca}
\eeq
where $C_C$ is the capture rate and $C_A$ is related to the annihilation rate $\Gamma_A$,
$\Gamma_A=C_AN_\chi^2$.
This has the solution
\beq
\Gamma_A={C_C\over 2} \tanh^2\left({t\over \tau}\right),
\eeq
where the equilibration time scale $\tau=1/\sqrt{C_CC_A}$. In most
cases for the Sun, and in the cases of observable fluxes for the
Earth, $\tau$ is much smaller than a few billion years, and
therefore equilibrium is often a good approximation ($\dot N_\chi=0$
in Eq.~(\ref{eq:ca})). This means that it is the capture rate
which is the important quantity that determines the neutrino flux.

The capture rate induced by scalar (spin-independent) interactions between
the neutralinos and the nuclei in the interior of the Earth or Sun is
the most difficult one to compute, since it depends sensitively on
Higgs mass, form factors, and other poorly known quantities. However,
this
spin-independent capture rate calculation is the same as for
direct detection treated in Section~\ref{subs:direct}. Therefore,
there is a strong correlation between the  neutrino flux expected from
the Earth (which is mainly composed of spin-less nuclei) and the signal
predicted in direct detection experiments \cite{begnu2,kamsad}. It
seems that even the large (kilometer-scale) neutrino telescopes
planned will not be competitive with the next generation of direct
detection experiments when it comes to detecting neutralino dark
matter,  searching for
annihilations from the Earth. However, the situation concerning
the Sun is more favourable. Due to the low counting rates for
the spin-dependent interactions in terrestrial detectors, high-energy
neutrinos
from the Sun constitute  a competitive and complementary neutralino dark matter
search. Of course, even if a neutralino is found through direct
detection, it will be extremely important to confirm its identity
and investigate its properties through indirect detection. In
particular, the mass can be determined with reasonable accuracy
by looking at the angular distribution of the detected
muons \cite{EG,BEK}.

For the  the Sun, dominated by hydrogen,
the axial (spin-dependent) cross section is  important and
relatively easy to compute. A good approximation
is given by \cite{jkg}
\beqa
     {C^{\rm sd}_\odot\over (1.3\cdot 10^{23}\, {\rm s}^{-1})} = &&\nonumber\\
\left({\rho_\chi\over 0.3\ {\rm GeV}\,{\rm cm}^{-3}}\right)
     \left({100\,{\rm GeV}\over m_\chi}\right)
\left({\sigma_{p\chi}^{\rm sd}\over 10^{-40}\ {\rm
     cm}^2}\right)
\left(270\ {\rm km/s}\over \bar v\right)
\eeqa
where $\sigma_{p\chi}^{\rm sd}$ is the cross section for
neutralino-proton elastic scattering via the axial-vector interaction,
$\bar v$ is the dark-matter
velocity dispersion, and
$\rho_\chi$ is the local dark matter mass.
The capture rate in the Earth is dominated
by scalar interactions, where there may be kinematic and other
enhancements, in particular if the mass of the neutralino almost
matches one of the heavy elements in the Earth. For this case,
a more detailed analysis is called for, but convenient approximations
are available \cite{jkg}. In fact, also for the Sun the
spin-dependent contribution can be important, in particular
iron may contribute non-negligibly.

\begin{figure}[!htb]\begin{center}
\epsfig{file=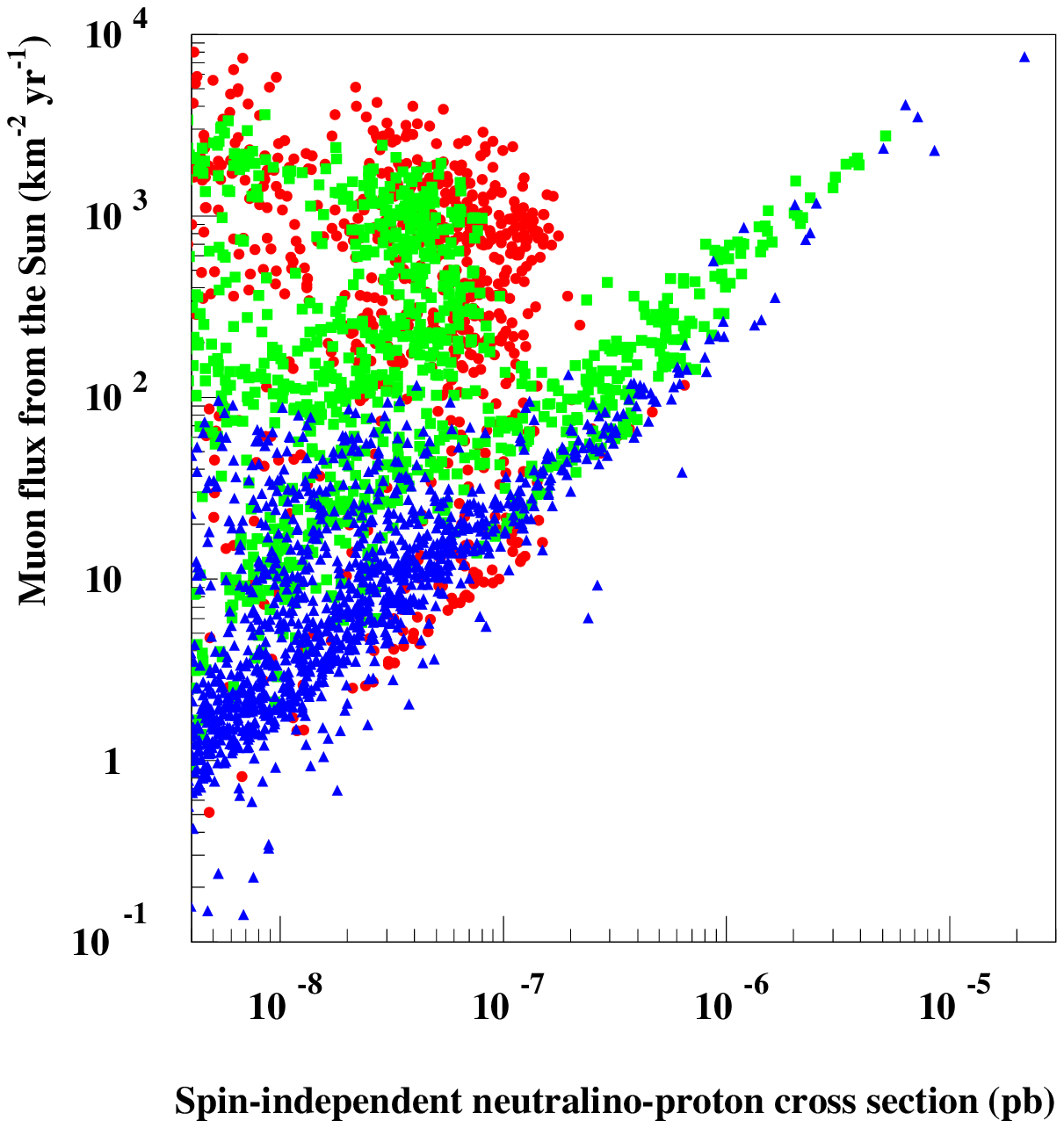,width=8cm}\epsfig{file=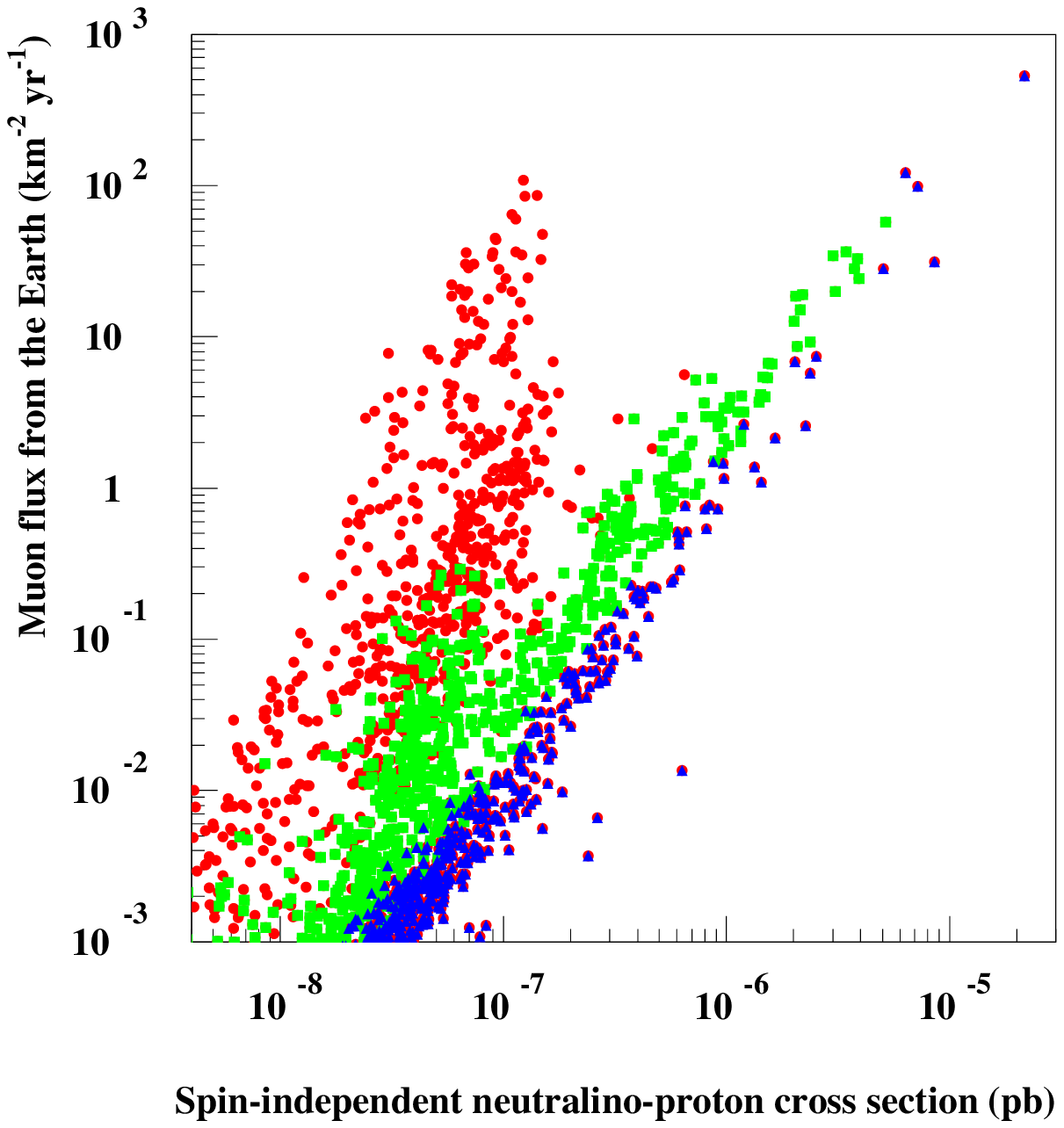,width=8cm}
\end{center}
         \caption{The predicted muon rates from
  neutralino annihilations in the Sun  (left-hand figure)
and in the Earth (right-hand figure) versus the
  neutralino-proton scattering cross section. A muon detection
threshold of 1 GeV has been assumed.
The requirements $m_{\chi^\pm}>95$ GeV, $m_{H_2}>100$ GeV,
 $0.1 < \Omega_\chi h^2 <0.2$ have been imposed.
For  details on the computational procedure, see \protect\cite{dkpop,begnu2}.
The filled circles denote higgsino-like 
models with gaugino fraction $Z_g < 0.01$,
squares are mixed models with  $0.01 < Z_g < 0.99$ and triangles are
gaugino  models
with $Z_g > 0.99$.}
         \label{fig:sunearth}
\end{figure}

To illustrate the potential of neutrino telescopes for discovery of dark matter
through neutrinos from the Earth or Sun, we present the results of a
full calculation \cite{begnu2} in Fig.~\ref{fig:sunearth}. 
The present experimental upper limits are of the order of a few 
thousand muon events per square kilometer and year (both for the
Sun and the Earth), and the irreducible background from the 
cosmic rays producing neutrinos in the Sun's atmosphere is of
the order of 20 (both the current limits and this background
are only wekly dependent on neutralino mass, see \cite{begnu2}).

In Fig.~\ref{fig:sunearth} it can be seen that a neutrino
telescope of area around 1 km$^2$, which is a size currently being discussed,
would have discovery potential for a range of  supersymmetric models.

It has recently been realized that even a small deviation from the
usually assumed Maxwellian velocity distribution of neutralinos in
the local part of the halo can have a large effect on the indirect
detection rates from the Earth \cite{dkpop}. In particular,
neutralinos which have scattered once by the Sun (going in
Keplerian orbits through the outer layers of the Sun) may have
their orbits perturbed by the large planets and will make up a new
population in phase space with velocities close to the Earth´s
orbital velocity around the Sun (but mainly going in nearly radial
orbits). This low-velocity population of neutralinos is very
efficiently captured by the Earth if $m_\chi\lsim 160 $ GeV, and
could give an enhanced muon flux by an order of magnitude.
This effect is included in Fig.~\ref{fig:sunearth}. The
correspondingly increased scattering rate in direct detectors is
not easily detected because of the low recoil energy, but a
direction-sensitive detector could be of great help \cite{dk2}.

If a signal were established, one can use the angular spread caused by
the radial distribution of neutralinos (in the Earth) and by the
energy-dependent mismatch between the direction of the muon and that
of the neutrino (for both the Sun and the Earth) to get a rather good
estimate of the neutralino mass \cite{EG}. If muon energy can also be
measured, one can do even better \cite{BEK}.

\section{Conclusions and outlook}\label{sec:conclusions}
To conclude, we have seen that the existence of dark matter
is more needed than ever, in order to explain a wealth
of new observations. In addition to
a large density of matter of unknown composition, the universe
seems to contain also a large quantity of dark energy, of
even more mysterious origin. Since the success of big bang
nucleosynthesis combined with the measured intensity
of the microwave background greatly constrains the amount of baryonic
matter allowed, the major part of the matter density
seems to be of non-baryonic origin.

The standard big bang model contains the required mechanism
to create a relic density of electrically neutral, stable
particles which, if their interactions are of electroweak
strength, may be perfect candidates for dark matter. Foremost
of these candidates are the hypothetical supersymmetric
partners of electrically neutral ordinary particles, motivated by
current thinking in particle physics. Also the more weakly
interacting low-mass axions which were invented to solve
the strong CP problem are attractive dark matter candidates,
probed by sensitive on-going experiments.

 Both  direct and indirect detection methods have the potential
of investigating supersymmetric and similar massive particle  dark
matter candidates. In particular, a combination of accelerator searches,
new solid state, liquid or gas detectors,
space-borne gamma-ray and air Cherenkov telescopes as well as neutrino telescopes
may have the sensitivity needed to rule out or confirm the supersymmetry
solution of the dark matter problem.

Since also the ex\-perimental situa\-tion con\-cerning massive neu\-trinos and
axions is getting clearer, there is a chance to reach the goal of
explaining the nature of the dark matter in the not too distant
future.

However, rapid success is by no means guaranteed. It may be that
the dark matter problem will plague the scientific community
for a long time also in the new millennium. With the realization that
dark vacuum energy may be an additional important component of our
Universe, the mystery of its inner workings has deepened. However,
this is a situation that is likely to inspire a young generation
of physicists and astronomers to even more spectacular progress
than the remarkable achievements since the 1930's when Zwicky
first took notice of the dark matter problem.
\ack
I wish to thank my collaborators, in particular Joakim Edsj\"o, Paolo Gondolo
and
Piero Ullio, for many
helpful discussions. Also, I want to thank Elliott Bloom, Sandro Bottino,
Jim Buckley, Per Carlson, Thibault
Damour, Ulf Danielsson, Nicolao Fornengo, Tom Francke, Tom Gaisser,
Ariel Goobar, Per Olof
Hulth, Marc Kamionkowski, Lawrence Krauss, Joel Primack,
Georg Raffelt, Hector Rubinstein, Pierre Sikivie, Joe Silk,
and Michael Turner for useful input to this review.

This work has been supported in part by the
Swedish Natural Science Research Council (NFR).

\end{document}